\documentclass[%
 aip,
 jmp,%
 amsmath,amssymb,
 reprint,%
]{revtex4-2}

\include{commands}
\usepackage{graphicx}
\usepackage{bm}
\usepackage{natbib}
\usepackage{caption}
\usepackage{subcaption}
\usepackage{color}
\usepackage{amsmath}
\usepackage{ulem}
\usepackage{xcolor,cancel}
\usepackage{tikz,siunitx}
\newsavebox{\astrutbox}
\sbox{\astrutbox}{\rule[-5pt]{0pt}{20pt}}

\begin{document}

\preprint{\it Physical Review Fluids}

\title[\it Dynamics of an elastoviscoplastic droplet in simple shear]{Dynamics of an elastoviscoplastic droplet in a Newtonian medium under shear flow}

\author{D. Izbassarov and O. Tammisola}
\email{outi@mech.kth.se} 
% NOTE: A full address must be provided: department, university/institution, town/city, zipcode/postcode, country.
\affiliation{$^1$FLOW, Engineering Mechanics, KTH Royal Institute of Technology, SE-10044 Stockholm, Sweden}

\date{\today}

\begin{abstract}
The dynamics of a single elastoviscoplastic drop immersed in plane shear flow of a Newtonian fluid is studied  by three-dimensional direct numerical simulations using a finite-difference/level set method combined with the Saramito model for the elastoviscoplastic fluid. This model gives rise to a yield stress behavior, where the unyielded state of the material is described as a Kelvin-Voigt viscoelastic solid and the yielded state as a viscoelastic Oldroyd-B fluid. Yielding of an initially solid drop of Carbopol is simulated under successively increasing shear rates. We proceed to examine the roles of nondimensional parameters on the yielding process; in particular, the Bingham number, the capillary number, the Weissenberg number and the ratio of solvent and total drop viscosity are varied. We find that all of these parameters have a significant influence on the drop dynamics, and not only the Bingham number. Numerical simulations predict that the volume of the unyielded region inside the droplet increases with the Bingham number and the Weissenberg number, while it decreases with the capillary number at low Weissenberg and Bingham numbers. A new regime map is obtained for the prediction of the yielded, unyielded and partly yielded modes as a function of the Bingham and Weissenberg numbers. The drop deformation is studied and explained by examining the stresses in the vicinity of the drop interface. The deformation has a complex dependence on the Bingham and Weissenberg numbers. At low Bingham numbers, the droplet deformation shows a non-monotonic behaviour with an increasing drop viscoelasticity. In contrast, at moderate and high Bingham numbers, droplet deformation always increases with drop viscoelasticity. Moreover, it is found that the deformation increases with the capillary number and with the solvent to total drop viscosity ratio. A simple ordinary differential equation model is developed to explain the various behaviours observed numerically. The presented results are in contrast with the heuristic idea that viscoelasticity in the dispersed phase always inhibits deformation. 
%The opposite trend can be explained by the properties of the Kelvin-Voigt viscoelastic model is described by a spring and dashpot mechanical system
%with the dashpot viscosity directly related to the solvent viscosity ratio and the spring elasticity inversely to the Weissenberg number.
\end{abstract}

\maketitle
\section{Introduction \label{sec:intro}} 
This manuscript considers the time-dependent, three-dimensional dynamics of an elastoviscoplastic (non-Newtonian) droplet surrounded by a Newtonian fluid in a simple shear flow. The dynamics of droplets in shear flows have attracted significant interest, mainly because this flow case governs the behaviour of dilute emulsions. Emulsions in turn play a pivotal role in a variety of applications, for example in food industry, chemical processing and biological materials. In order to control these industrial processes, it is hence important to understand the dynamics of a droplet in shear flow \citep{Windhab2005}. However, most studies so far are restricted to the case where both the droplet and the other fluid are Newtonian. In real life, both components are often non-Newtonian. It has been shown that non-Newtonian fluids exhibit exotic behaviors which significantly affect the drop deformation and breakup. This study considers a droplet of a particular type of non-Newtonian fluid, which exhibits yield-stress (plasticity) along with elasticity. Droplets of such fluids have not been numerically simulated before, to the authors' knowledge. In simple shear flow, no results for droplets of any kind of yield-stress fluids (including purely viscoplastic) could be found in the literature; our results at low Weissenberg numbers may shed light on the drop behaviour in the viscoplastic limit. 

Yield-stress fluids behave like solids below a local stress threshold (the yield stress), and flow like liquids above this threshold \citep{BalmforthAnnuRev}. Yield-stress fluids can be found in geophysical applications, such as mudslides and the tectonic dynamic of the Earth. They are also found in industrial applications such as mining operations, the conversion of biomass into fuel, and the petroleum industry, to name a few. Biological and smart materials can have a yield stress, making yield-stress fluid flows relevant for problems in physiology, biolocomotion, tissue engineering, and beyond. Most of these applications deal with multiphase flows or interfacial flows (\citet{maleki2018submergedjet, chaparian2017cloaking, maleki2015macro} among many others). Therefore, there is a compelling need to study droplet-laden flows with yield stress and predict their flow dynamics in various situations, including three-dimensional and inertial flows.

The simplest physical model of a yield-stress fluid assumes that the material is a rigid solid at stresses lower than the yield stress. This assumption leads to purely viscoplastic models, such as the Bingham model \citep{bingham1922fluidity}, where the yielded fluid flow is Newtonian, and the Herschel-Bulkley model \citep{herschel1926measurement}, where the yielded fluid flow is shear-thinning. However, many yield-stress fluids, of which Carbopol is one example, deform like elastic solids in the unyielded state and behave as viscoelastic liquids in the yielded state, displaying elastic (E), viscous (V) and plastic (P) properties. 
Numerical simulations of EVP fluid flows are not a straightforward task due to the inherent nonlinearity of the governing equations. Nevertheless, numerical simulations are needed to provide quantitative information that is extremely difficult to access experimentally in yield-stress fluids (for example, the yielded and unyielded regions, and velocity fields and stress fields separated into different contributions),
and also understanding the physics of the interaction between droplets and the surrounding fluid.

Numerical simulations have already helped to reveal elastic effects in yield-stress fluid flows around particles, in liquid foams and Carbopol. First, \citet{dollet2007two} performed experimental measurements for the flow of liquid foam around a circular obstacle, where they observed an overshoot of the velocity (the so-called negative wake) behind the obstacle. Then, \citet{cheddadi2011understanding} simulated
the flow of an EVP fluid around a circular obstacle by employing Saramito's EVP model \citep{saramito2007new}. The numerical simulation using
the EVP model captured the negative wake. A purely viscoplastic flow model (Bingham model) on the other hand always
predicted fore-aft symmetry and the lack of a negative wake, in contrast with the aforementioned experimental observations.
The numerical simulations could hence prove that the negative wake was the consequence of the foam elasticity. Recently, the loss of the
fore-aft symmetry and the formation of the negative wake around a single particle sedimenting in a Carbopol solution was captured by transient numerical calculations by \citet{Fraggedakis_SM_2016} by adopting the Saramito EVP model, in agreement with the experimental observations in Carbopol gel \citep{holenberg2012experiment}. The elastic effects on viscoplastic fluid flows have also been addressed in numerical simulations of the EVP fluids through an axisymmetric expansion-contraction geometry \citep{nassar2011flow}. It was observed
that elasticity alters the shape and the position of the yield surface remarkably, and elasticity needs to be included to reach
qualitative agreement with experimental observations for the flow of Carbopol aqueous solutions. More recently, elastoviscoplastic effects have been analysed in porous media flow \citep{DEVITA201810} and in turbulent channel flow \citep{EVPturbulence1}.  

Regarding droplets in non-Newtonian fluids, a large body of literature has been devoted to viscoelastic fluids. Here, we only review studies on droplets in low Reynolds number viscoelastic shear flows for conciseness, although it needs to be mentioned that elastic effects (loss of fore-aft symmetry at low Reynolds) are commonly seen for buoyancy-driven bubbles and droplets \citep{Pillapakkam_JFM_2007, Fraggedakis_JFM_2016}. In simple shear flows, viscoelasticity in the drop tends to reduce drop deformation. The steady state deformation of an Oldroyd-B viscoelastic droplet under shear in Newtonian fluid was considered in several works \citep{Yue_JFM_2005,Aggarwal_JFM_2007,Aggarwal_JFM_2008}. With increasing elasticity (increasing Weissenberg number), the droplet deformed less, while the inclination angle of the droplet increased. Similar behaviour was found by \citet{Mukherjee_JNNFM_2009} in viscosity-matched systems. However, when the droplet viscosity was higher than that of the surrounding fluid, they found that the droplet deformation started to increase with elasticity, at high Weissenberg numbers. The stresses on a single droplet were analysed and generalized towards stresses in dilute emulsions by \citet{Aggarwal_JNNFM_2008}. It was found that the effective shear viscosity of the emulsion was not much affected by droplet phase viscoelasticity, while the normal stress differences increased with drop viscoelasticity at high capillary numbers. The effect of confinement of the viscoelastic droplet in microfluidic applications has also been analysed \citep{Guido2011,Cardinaels_JNNFM_2011}, whereby it was found that confinement increased the droplet deformation and promoted breakup, and the viscoelastic stresses also increased.  

While droplet viscoelasticity often just reduces deformation at the steady state, the time-dependent dynamics may change qualitatively. A new breakup mode due to droplet viscoelasticity was found by \citet{Verhulst_JNNFM_2009part2}. The effect of maximal extensibility of the polymers (using the FENE-P numerical model) was studied by \citet{Gupta2014}, who found a nontrivial relation between the polymer extensibility and the droplet breakup behaviour. In systems where the droplet viscosity was much higher than that of the surrounding fluid, the droplet was seen to oscillate periodically \citep{Mukherjee_JNNFM_2009}.

Problems where the surrounding fluid (matrix) is viscoelastic but the droplet is Newtonian have also been analysed. As the present work deals with the opposite configuration, we will not attempt to review these results comprehensively. However, \citet{Verhulst_JNNFM_2009} compared the effects of droplet viscoelasticity and matrix viscoelasticity experimentally and computationally, and found that the matrix viscoelasticity generally has a much larger effect on the droplet deformation and orientation. Early works on the effect of Weissenberg (or Deborah) number gave contradictory results: in some experiments, the droplet deformation decreased with matrix elasticity \citep{Flumerfelt_1972}, while other experiments found that the deformation increased \citep{Elmendorp_1985, Mighri_1998}. The main reason for this apparent contradiction is that matrix viscoelasticity has a non-monotoneous effect when the elasticity is increased: it reduces the droplet deformation at low Weissenberg numbers (weak viscoelasticity), while it increases the deformation at high Weissenberg numbers (strong viscoelasticity). The non-monotoneous dependence on the Weissenberg number was first discovered in the computational study of \citet{Yue_JFM_2005}, and later in a fully three-dimensional simulation by \citet{Aggarwal_JFM_2008}. 

The present work investigates how the dynamics of the droplet in shear flow change, when the droplet is elastoviscoplastic rather than viscoelastic. The EVP behavior is obtained from the Saramito \citep{saramito2007new} constitutive equation, without shear-thinning or time-dependent rheology effects. The manuscript is organized as follows. The governing equations and their numerical solution are described in Section~\ref{sec:formulation}. In Section~\ref{sec:problem}, the flow case and its characteristic nondimensional parameters are presented.  Numerical results are shown in Section~\ref{sec:yielding} for varying Bingham, Weissenberg and capillary numbers, and solvent-to-total viscosity ratio. We consider the distribution of yielded and unyielded regions inside the droplet. Next in Section~\ref{sec:comparison}, we look at how the droplet deformation and inclination angle change when the above parameters are varied, and explain the change from a simple ordinary differential equation (ODE) model for fully unyielded droplets. Finally, we study the stress distributions along the interface (Section~\ref{sec:emulsions}). The work is concluded in Section~\ref{sec:concl}. Finally, the formulation of the ODE model is included as an Appendix~\ref{sec:ODE}. 

\section{Governing equations and their numerical solution \label{sec:formulation}}

The governing equations for this work are the momentum balance equations for two incompressible, immiscible fluids, combined with the constitutive equation for the elastoviscoplastic stresses. The numerical method that allows 3D numerical simulations of these equations is described in \citet{Izbassarov_IJNMF_2018}. At the interface between the two fluids, velocity and tangential stress shall remain continuous, while there is a jump in the normal stress due to surface tension. These conditions were explicitly formulated in \citet{Tammisola_JFM2012b}, where a two-domain approach was used. In the present work however, the coupling terms between the two fluids (droplet and matrix) are treated using the level set method \citep{Sussman_JCP_1994}. Hence, we adopt a continuous formulation of the governing equations \citep{Tammisola_PRF_2017,Izbassarov_IJNMF_2018}, with a delta function representing the normal stress jump due to surface tension. The level-set method used here is similar to the one used in our other works with droplets in low Reynolds number flows \citep{Ge_AX_2017, Ge_SoftMatter}, with the difference that the ghost-fluid method (GFM) is here replaced by a more diffusive continuum surface force (CSF) method \citep{}, to simplify the treatment of the non-Newtonian stresses. 

In the following, all dimensional quantities are denoted by stars. The governing equations for the dynamics of an incompressible flow of two immiscible fluids can be written as:
\begin{subequations}
 \begin{equation}
   \nabla \cdot {\bf u}^* = 0,
  \label{div free}
 \end{equation}
 \begin{equation}
   \rho^* \bigg( \frac{\partial \bf{u}^*}{\partial t^*} + {\bf u}^* \cdot \nabla {\bf u}^* \bigg) = -\nabla p^* 
+\nabla \cdot \mu_s^*  ( \nabla {\bf u}^* + \nabla {\bf u^*}^{T} ) + \nabla \cdot \bm{\tau}^* + \Gamma^* \kappa^* \delta(\phi^*) \bf n^*,
  \label{NS}
 \end{equation}
\label{mainNS}
\end{subequations}

\noindent where ${\bf u}^* = {\bf u}^* \left( {\bf x}^*, t^* \right)$ is the velocity field, $p^* = p^* \left( {\bf x}^*, t^* \right)$ the pressure field and ${\bm \tau}^* = {\bm \tau}^* \left( {\bf x}^*, t^* \right)$ an extra stress tensor defined below. The function $\phi^*({\bf x}^*,t^*)$ is the level-set function approximating the signed distance from the interface. Hence, $\phi^* = 0$ denotes the interface, $\phi^* > 0$ denotes fluid 1 and $\phi^* < 0$  fluid 2. The last term in Eq. (\ref{NS}) is a body force due to surface tension, where ${\bf n}^*$ is the unit normal vector to the interface, $\kappa^*$ the local mean curvature, $\delta$ a regularized delta function and $\Gamma^*$ the surface tension. Finally, $\rho^*$ and $\mu_s^*$ are the density and solvent viscosity of the fluid. 

In the present study, the elastoviscoplastic (EVP) effects in the flow are reproduced by the extra stress tensor $\bm{\tau}^*$, defined with the elastoviscoplastic Saramito model as 
\begin{equation}
\lambda^* \bigg(\frac{\partial \bm{\tau}^*}{\partial t^*}+{\bf u}^*\cdot \nabla \bm{\tau}^*-\bm{\tau}^*\cdot \nabla {\bf u}^*-\nabla {\bf u^*}^{T}\cdot \bm{\tau}^* \bigg)
+{\rm{max}}\left[0,1-\frac{\tau_0^*}{|\bm{\tau_{d}}^*|}\right]\bm{\tau}^* =\mu_p^*(\nabla {\bf u}^* + \nabla {\bf u^*}^{T} ), 
\label{SRM}
\end{equation}

\noindent where $\lambda^*$ is the relaxation time, $\mu_p^*$ is polymeric viscosity, $\tau_0^*$ is the yield stress, $\bm{\tau_{d}}^*$ is the deviatoric part of $\bm{\tau}^*$ and $|\cdot|$ denotes the usual Euclidean matrix norm. The second deviatoric stress tensor $\boldsymbol{\tau_d}^*$ and its magnitude is defined as in previous works\cite{Fraggedakis_SM_2016, DEVITA201810, EVPturbulence1, Izbassarov_IJNMF_2018}: $|\boldsymbol{\tau}_d|= \sqrt{\frac{1}{2} \tau_{d,ij}\tau_{d,ij}}$. We can observe that for $\tau_0^*=0$, the Oldroyd-B polymeric liquid model is recovered, and that if yield stress is large ($\tau_0^* \to \infty$), the equation describes a viscoelastic solid. The density, the solvent and the polymeric viscosities and the relaxation time vary across the fluid interface and are expressed in a continuous form as
 \begin{equation}
 \begin{split}
  \rho^* &= \rho_1^*H(\phi) + \rho_2^* \big(1-H(\phi)\big), \quad
  \lambda^* =  \lambda_1^*H(\phi) + \lambda_2^* \big(1-H(\phi)\big), \\
  \mu_s^* &=  \mu_{s,1}^*H(\phi) +  \mu_{s,2}^*\big(1-H(\phi)\big), \quad
  \mu_p^* = \mu_{p,1}^*H(\phi) +  \mu_{p,2}^*\big(1-H(\phi)\big),
 \end{split}
 \end{equation}
 \label{mat prop}

\noindent where $\phi$ is the level set function denoting the distance to the interface, the subscripts $1$ and $2$ denote the properties of the bulk and drop fluids, respectively, and $H(\phi)$ is the Heaviside function.

The flow equations (\ref{div free} \& \ref{NS}) are solved in parallel, fully coupled with the EVP model equations (\ref{SRM}), by the numerical method recently presented in \citet{Izbassarov_IJNMF_2018}. The flow and the EVP model equations are solved on a staggered uniform Cartesian grid using a projection method. The spatial derivatives are approximated using second-order central differences, apart from the advection terms in the EVP constitutive, level-set advection and reinitialization equations, which are discretized by the fifth-order weighted essentially non-oscillatory (WENO) scheme \citep{Liu_JCP_1994}. The time integration is performed by the second-order Adams-Bashforth scheme (AB2) for both flow and elastoviscoplastic model equations. The time-integration in the level set advection and reinitialization equations is performed using a three-stage total-variation-diminishing (TVD)
third-order Runge-Kutta scheme \citep{Shu_JCP_1988}. A complete description of the level-set method can be found in \citet{Sussman_JCP_1994} and \citet{Ge_AX_2017}, and the treatment of the EVP multiphase flow in \citet{Izbassarov_IJNMF_2018}. 

\begin{figure}
\begin{center}
\begin{tabular}{cc}
{\includegraphics[width=0.7\textwidth]{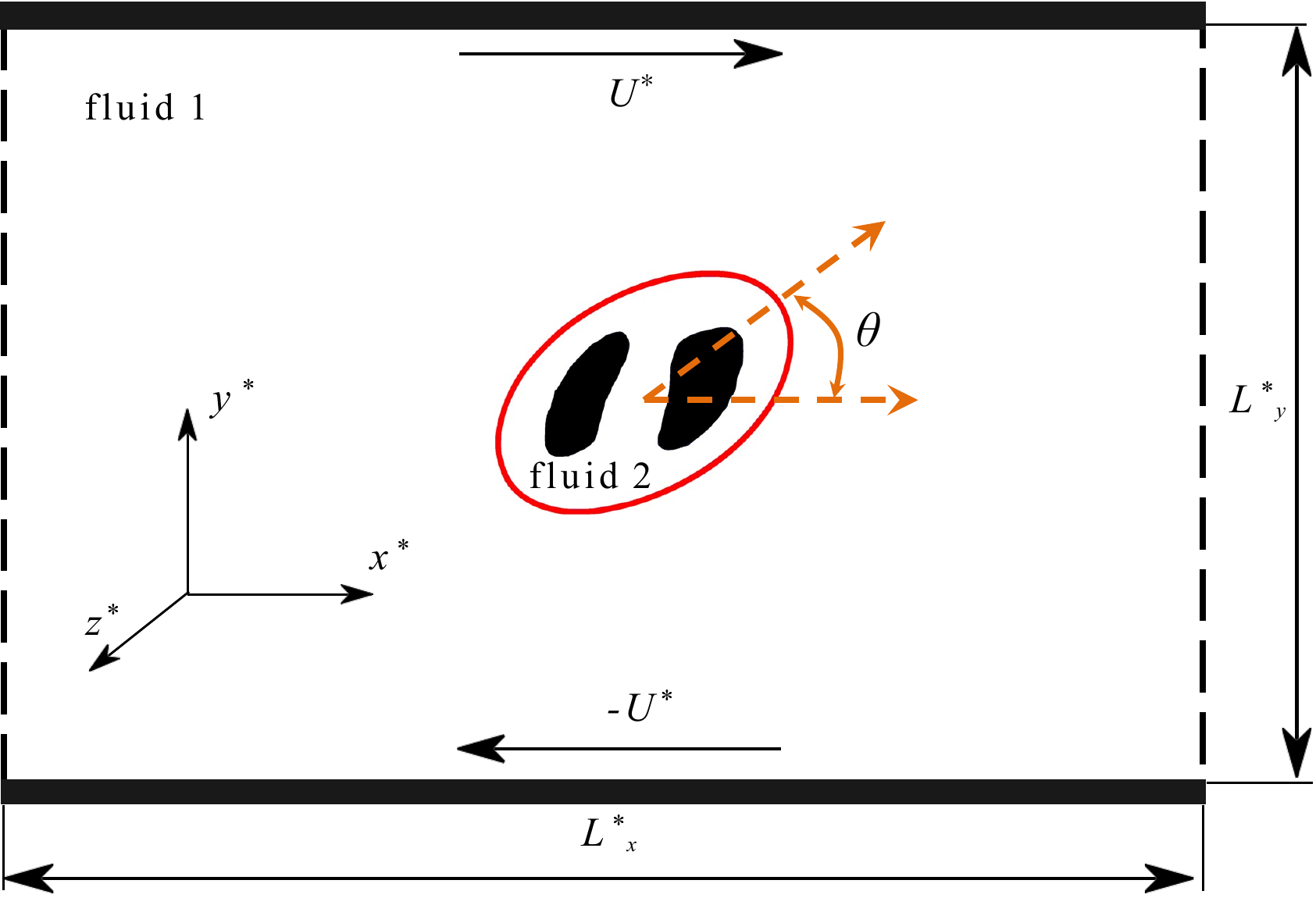}}
\end{tabular}
\end{center}
\caption{Geometry of the shear-driven system with an elastoviscoplastic droplet.}
\label{sdgeom}
\end{figure}

\section{Problem statement \label{sec:problem}}

In the present work, we consider a three-dimensional elastoviscoplastic (EVP) droplet in a simple shear flow of a Newtonian liquid. The configuration is sketched in figure \ref{sdgeom}, and $x^*$ is the streamwise, $y^*$ the vertical and $z^*$ the spanwise coordinate. The droplet is initially spherical and fully solid with a radius $R^*$. Velocities $U^*$ and $-U^*$ in the streamwise $x^*-$direction are imposed at the upper and lower walls, respectively, to obtain a shear rate $\dot{\gamma}^*=2U^*/L_y^*$, which deforms the droplet. Under shear flow, the droplet may yield partly or fully, \textit{i.e.} go from solid to liquid. The solid region is represented by the black colour in the following. As boundary conditions, we impose no-slip at the walls, and periodic boundary conditions in the streamwise and spanwise directions. The channel height is $L_y^*=8R^*$, resulting in a weak confinement of the droplet.

The lengths are nondimensionalized with $R^*$, and velocities with $\dot{\gamma}^* R^*$ (whereby the shear flow time scale $\dot{\gamma^*}^{-1}$ has been chosen). The stresses and the pressure are nondimensionalized by the viscous stress scale of the outer flow $\mu_1^* \dot{\gamma}^*$. The outer fluid (\textit{i.e.} matrix) density $\rho_1^*$ and dynamic viscosity $\mu_1^*$ are taken as the reference values for nondimensionalisation. The problem is characterized by six non-dimensional parameters. Firstly, we have the droplet Reynolds number $Re=\rho_1^* \dot{\gamma^*}{R^*}^{2}/\mu_1^*$, and the capillary number $Ca=R^*\dot{\gamma}^* \mu_1^*/\Gamma^*$ characterizing the importance of surface tension compared to viscous forces. Furthermore, because the droplet is elastoviscoplastic, the Bingham number $Bi=\tau_0^*/(\mu_1^* \dot{\gamma}^*)$ emerges as a ratio between the droplet yield stress $\tau_0^*$ and the matrix viscous stress. The elastic effects inside the droplet are characterized by the Weissenberg number $Wi=\lambda^* \dot{\gamma}^*$, where $\lambda^*$  is the elastic time scale. The last two parameters are the droplet-to-matrix viscosity ratio $k_\mu=\mu_2^*/\mu_1^*$ and the density ratio $k_\rho=\rho_2^*/\rho_1^*$. In the previous definitions, $\mu_2^*=\mu_{s,2}^*+\mu_{p,2}^*$ and $\mu_{s,2}^*$ are the total (sum of the solvent and polymeric viscosities) and the solvent viscosity of the fluid 2, respectively. The solvent viscosity ratio is defined as $\beta=\mu_{s,2}^*/\mu_2^*$. Considering that only the droplet (fluid 2) is EVP, the equations \ref{mainNS} \& \ref{SRM} can be rewritten in a non-dimensional form as 
\begin{subequations}
 \begin{equation}
   \nabla \cdot {\bf u} = 0,
  \label{div free1}
 \end{equation}
 \begin{equation}
   \rho Re \bigg( \frac{\partial \bf{u}}{\partial t} + {\bf u} \cdot \nabla {\bf u} \bigg) = -\nabla p
+\nabla \cdot \mu_s  ( \nabla {\bf u} + \nabla {\bf u}^T ) + \nabla \cdot \bm{\tau} + Ca^{-1} \kappa \delta(\phi) \bf n,
  \label{NS1}
 \end{equation}
\begin{equation}
Wi \bigg(\frac{\partial \bm{\tau}}{\partial t}+{\bf u}\cdot \nabla \bm{\tau}-\bm{\tau}\cdot \nabla {\bf u}-\nabla {\bf u}^T\cdot \bm{\tau} \bigg)
+{\rm{max}}\left[0,1-\frac{Bi}{|\bm{\tau_{d}}|}\right]\bm{\tau} =k_\mu (1-\beta)(\nabla {\bf u} + \nabla {\bf u}^T ),
\label{SRM2}
\end{equation}
\end{subequations} 
\noindent 
where $\rho$ and $\mu_s$ are the nondimensional counterparts of $\rho^*$ and $\mu_s^*$ in the previous section. To quantify the deformation of the droplet in the shear plane ($x-y$ plane), we use the Taylor deformation parameter $D=(L-B)/(L+B)$, where $L$ and $B$ are the major and minor axis of the equivalent ellipsoid in the middle plane. The inclination angle of the drop axis with respect to the streamwise direction is denoted by $\theta$. Using a methodology developed by \citet{Ramanujan_JFM_1998}, the equivalent ellipsoid is obtained by the inertia tensor of the drop.

\begin{figure}
\begin{center}
\begin{tabular}{cc}
{\includegraphics[width=0.45\textwidth]{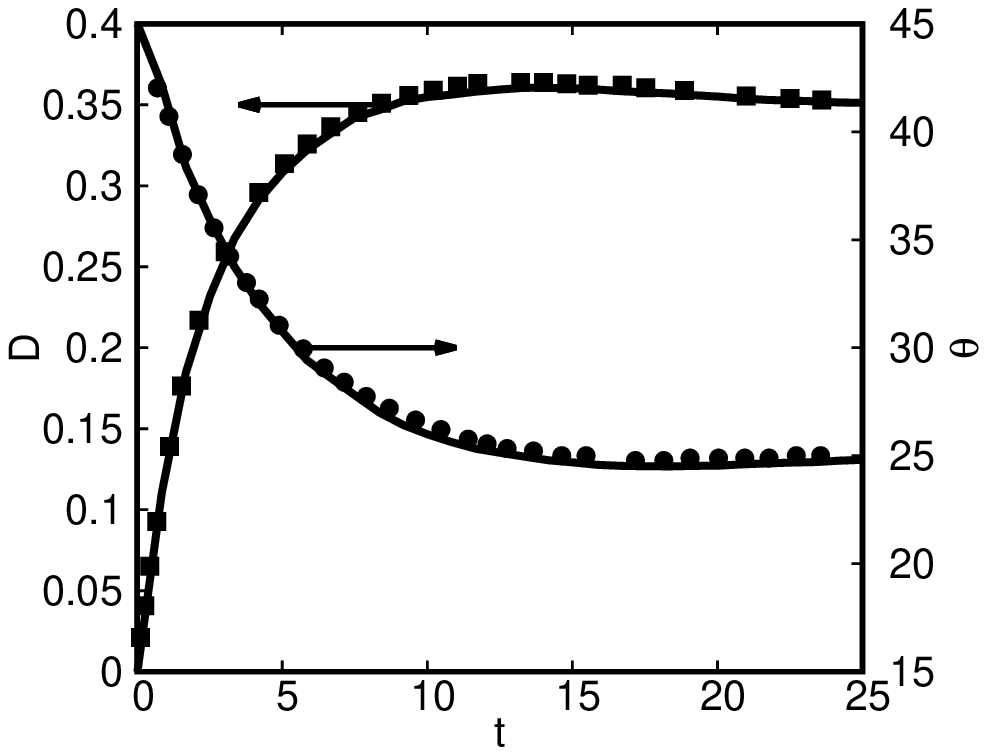}} 
{\includegraphics[width=0.45\textwidth]{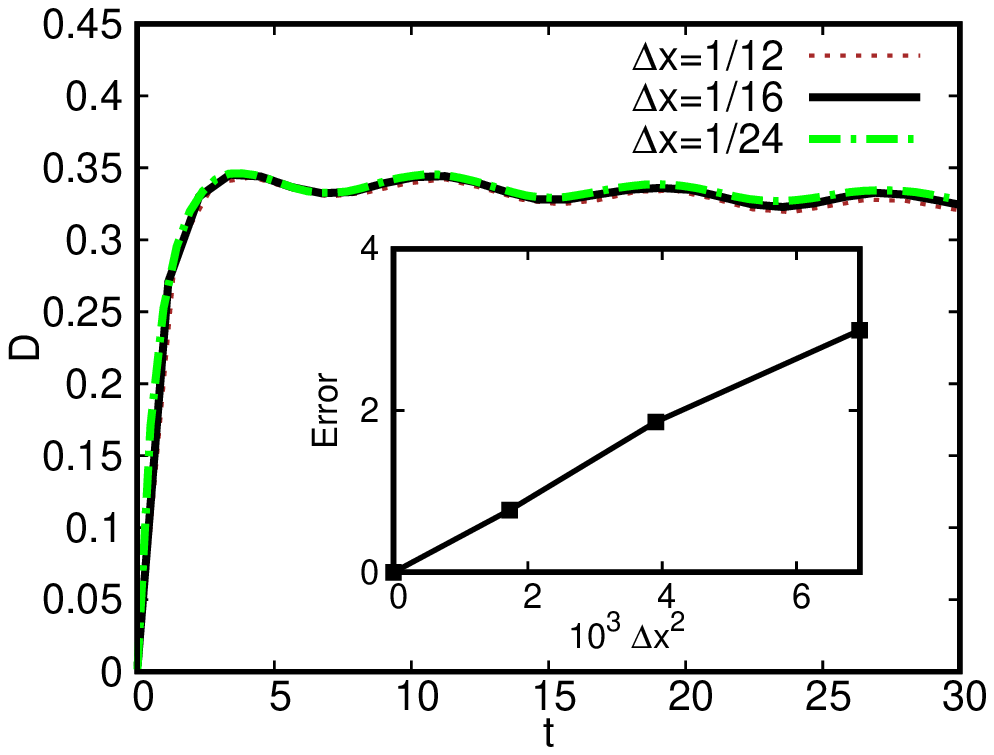}}
\end{tabular}
\end{center}
\caption{Left: Code validation. Time evolutions of the droplet deformation parameter $D$ and inclination angle $\theta$ in our simulations with $Bi=0$ (solid line) are compared
%and at $128 \times 256 \times 128$ resolution 
 against those of \citet{Verhulst_JNNFM_2009} (symbols). Right: Grid convergence study.  The deformation parameter $D(t)$ for $Bi=1.6$ at various grid resolutions with inset showing convergence in $\rm{Error}=100\times(\rm{D}_{\Delta x\rightarrow0}-\rm{D}_{\Delta x})/\rm{D}_{\Delta x\rightarrow0}$ with increasing grid resolution.}
\label{GConv}
\end{figure}

In the following, numerical simulations are performed to study the dynamics of EVP droplet in a Newtonian fluid under shear flow. The computational domain (see figure \ref{sdgeom}) has dimensions $L_x \times L_y \times L_z=16R \times 8R \times 8R$ and is resolved by a grid of
$\Delta x= \Delta y = \Delta z = R/16$ (32 points per drop diameter) in all the results presented in this paper. A grid convergence study has been performed (Fig.~\ref{GConv}) to ensure that the solutions are grid independent, i.e., the spatial error is below $3\%$. The solver has also been validated against the results of \citet{Verhulst_JNNFM_2009} for a viscoelastic droplet in a Newtonian shear flow (Fig.~\ref{GConv}), and several other validation viscoelastic and EVP validation cases can be found in \citet{Izbassarov_IJNMF_2018}. 

\section{Results on the EVP droplet yielding process \label{sec:yielding}}

\begin{table}
\caption{Specification of the dimensional parameters for Carbopol, based on Ref. \cite{Fraggedakis_SM_2016}}
\centering
\begin{tabular}{c c c c c}
  \hline
    $\mu^*_s (Pa s)$ & \quad  $\mu^*_p (Pa s)$ & \quad  $\lambda^* (s)$ & \quad  $\Gamma^* (N/m)$ & \quad $\tau_0^* (Pa)$\\
  \hline
   0.12 & \quad 2.23 & \quad 0.129 & \quad 0.034 & \quad 1.6\\
  \hline
\end{tabular}
\label{Material}
\end{table}
\begin{table}
\caption{Parameters for simulations at different shear rates for Carbopol droplet.}
\centering
\begin{tabular}{ c c c c c }
  \hline
  $\dot{\gamma}^* (s^{-1})$ & \quad  $Ca$ & \quad $Bi$ & \quad $\beta$ & \quad $Wi$ \\
   \hline
  0.32 & 0.1 & \quad 3.2  & \quad 0.05  & \quad  0.04\\
  0.43 & 0.13 & \quad 2.4  & \quad 0.05  & \quad  0.05\\
  0.64 & 0.2 & \quad 1.6  & \quad 0.05  & \quad  0.08\\
 \hline
\end{tabular}
\label{ND_par}
\end{table}

\begin{figure}
\begin{center}
\begin{tabular}{c}
{\includegraphics[width=\textwidth]{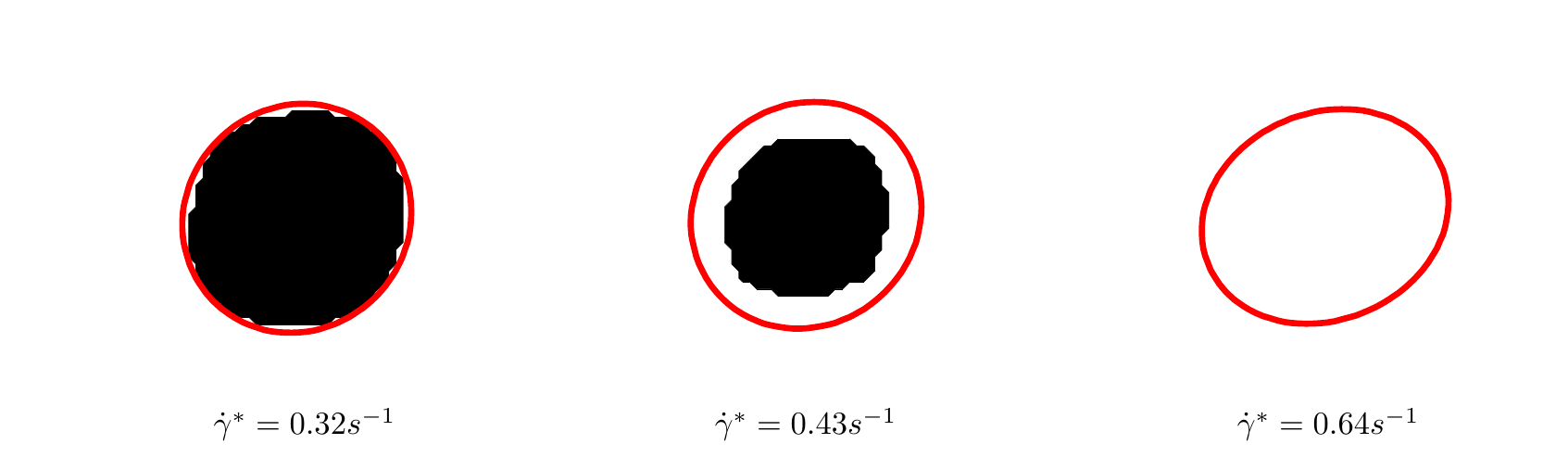}}
\end{tabular}
\end{center}
\caption{Effect of the shear rate on the Carbopol solution (Table~\ref{Material}).  The unyielded region is obtained from $\mathcal{F} \le 10^{-3}$,  where $\mathcal{F}=1-(Bi/|\bm{\tau_{d}}|)$.}
\label{SolidDim}
\end{figure}

\subsection{Droplet of a yield-stress fluid under shear - a numerical experiment \label{sec:numexp}}

We start the study by imitating an experiment, where a droplet of a given EVP material is placed in simple shear flow, under different levels of shear ( $\dot{\gamma^*}=U^*/H^*$). The EVP material properties are based on the data provided by \cite{Fraggedakis_SM_2016} and are given in Table \ref{Material} $\&$ \ref{ND_par}. It is same Carbopol solution, but all of its parameters vary with shear rate. Note that here we employed a low but numerically tractable value of $\beta$, while the actual value for Carbopol is lower. Also for numerical reasons, we increased the minimal Reynolds number to $Re=0.05$.

Figure~\ref{SolidDim} shows the droplet interface at steady state, together with the unyielded regions defined by $\mathcal{F}=10^{-3}$, where $\mathcal{F}=1-(Bi/|\bm{\tau_{d}}|)$,  at different $\dot{\gamma}^*=0.32, 0.43$ and $0.64 s^{-1}$. We have observed that the shape of the unyielded region is insensitive to smaller values of $\mathcal{F}$, and hence it is fixed for all results presented in this work. At the lowest shear rate, the droplet remains solid almost everywhere. Similarly to an experiment, while keeping the material constant, we now increase the shear rate. Increasing the shear rate changes its state from fully unyielded to partly yielded and finally to fully yielded (fluidized everywhere). This qualitative development is expected for any EVP droplet because its Bingham number decreases with shear; the shear stress imposed by the external shear flow becomes stronger than the droplet yield stress. 

In this dimensional experiment, $Wi$, $Bi$ and $Ca$ all change simultaneously, and therefore, the effects of each parameter on the yielding process are difficult to separate and understand.
In the next sections, we will adopt a nondimensional formulation and study the effects of each nondimensional parameter in detail.

% New figure 5: Time evolution of solid regions, low Wi
\begin{figure}
\begin{center}
\begin{tabular}{c}
{\includegraphics[width=\textwidth]{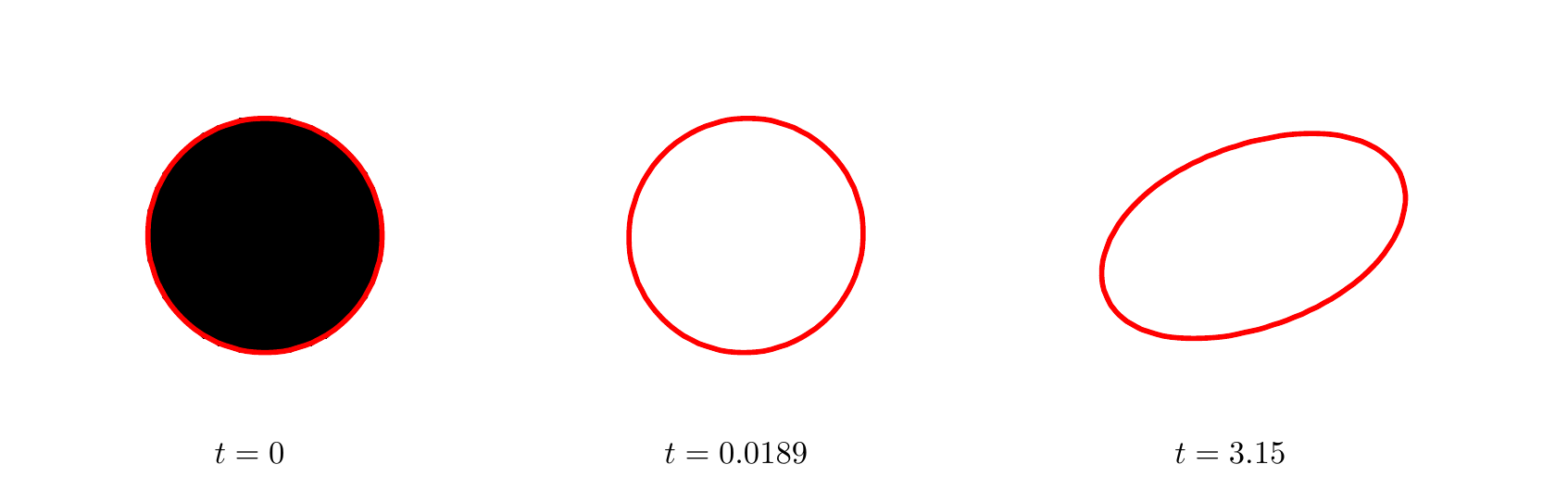}}
\end{tabular}
\end{center}
\caption{Time-evolution of unyielded (black) and yielded (white) regions inside the EVP droplet at $Bi=0.8$ and $Wi=0.01$. The unyielded region is obtained from $\mathcal{F} \le 10^{-3}$,  where $\mathcal{F}=1-(Bi/|\bm{\tau_{d}}|)$.}
\label{Solid0p25_0p01}
\end{figure}

% New figure 6: Time evolution of solid regions, medium Wi
\begin{figure}
\begin{center}
\begin{tabular}{c}
{\includegraphics[width=\textwidth]{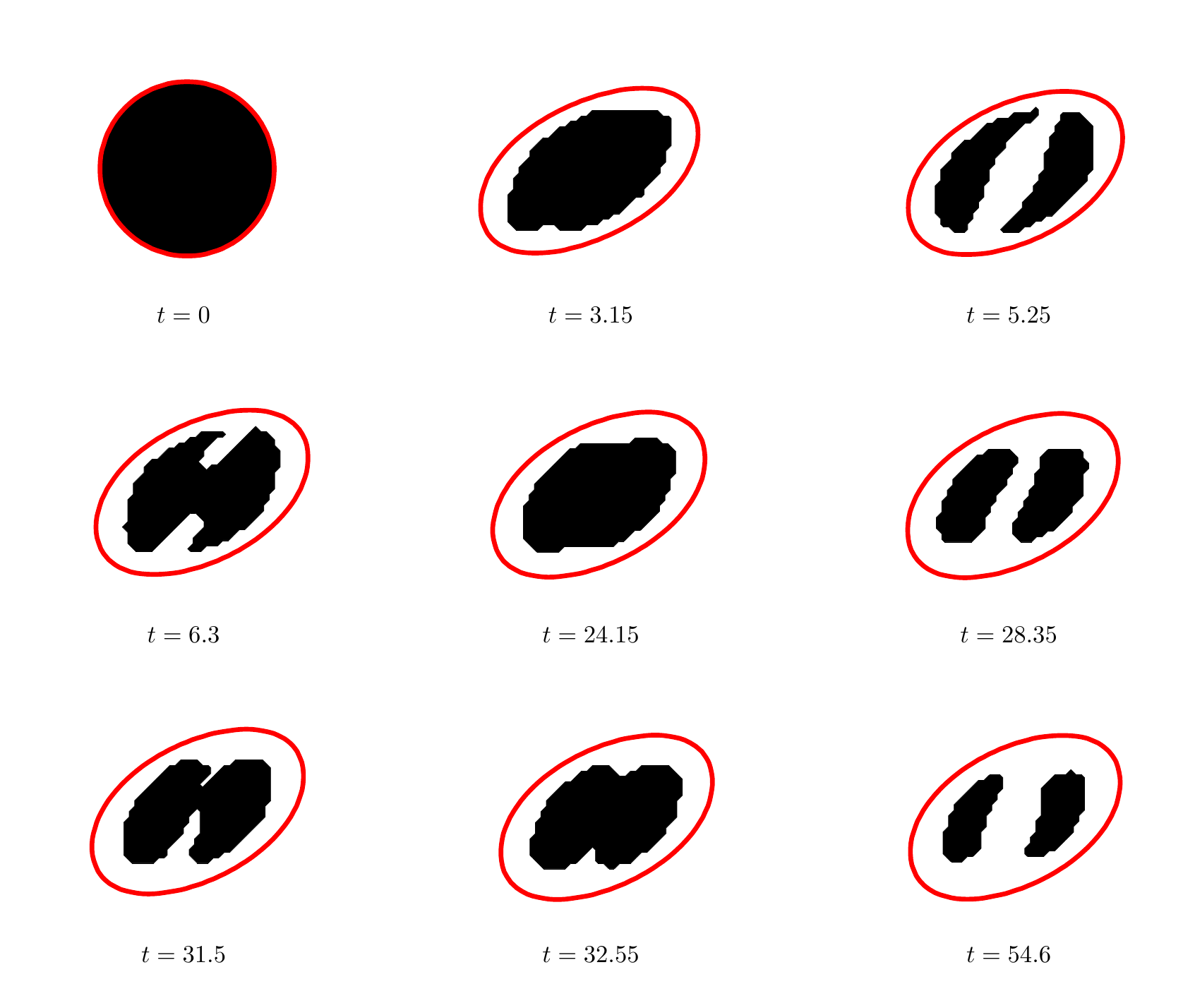}}
\end{tabular}
\end{center}
\caption{Time-evolution of unyielded and yielded regions inside the EVP droplet at $Bi=0.8$ and $Wi=0.5$. The unyielded region is represented by the black colour and obtained from $\mathcal{F} \le 10^{-3}$,  where $\mathcal{F}=1-(Bi/|\bm{\tau_{d}}|)$.}
\label{Solid0p25_0p5}
\end{figure}

%New figure 7: Velocity distributions at steady state
\begin{figure}
\begin{center}
\begin{tabular}{c}
{\includegraphics[width=\textwidth]{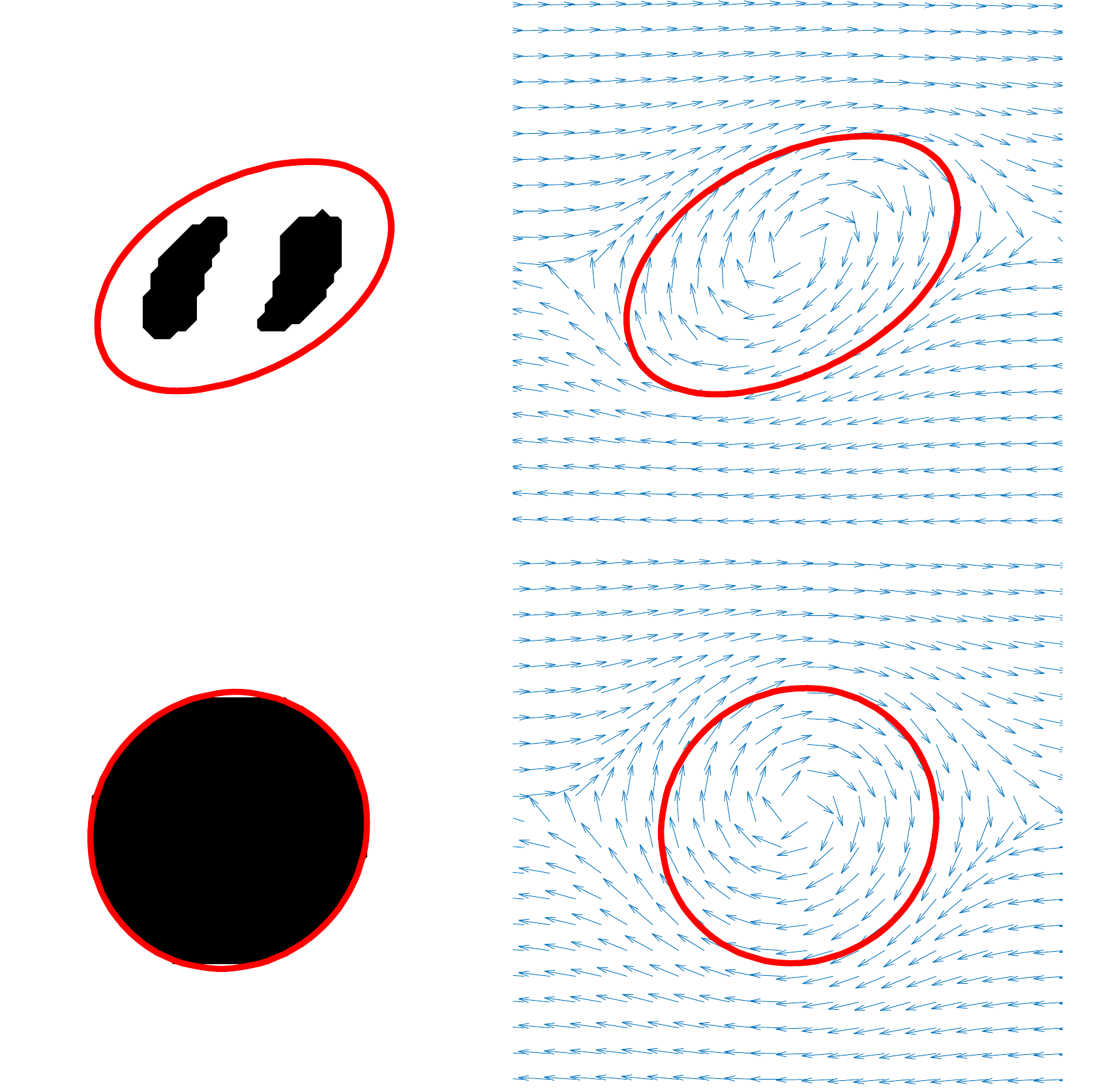}}
\end{tabular}
\end{center}
\caption{Top row: EVP droplet at $Bi=0.8$ and $Wi=0.5$, unyielded regions (left) and velocity vectors (right). Bottom row: EVP droplet at $Bi=8$ and $Wi=0.01$, unyielded region (left), and velocity vectors (right).}
\label{StreamBi}
\end{figure}

\subsection{Time evolution of the yielding process}

We will start by characterizing the flow in and around an EVP droplet in time. Following Ref.~\cite{Verhulst_JNNFM_2009},\footnote{Note that this is a value of a polymeric fluid, while lower $\beta$ are typical for Carbopol. While individual results may be affected by the value of $\beta$, we have verified that all trends that we describe are independent of the choice of $\beta$.}, the base case parameters are chosen as $Re=0.05, Ca=0.32, Wi=2.31, \beta=0.68, k_{\rho}=1$ and $k_{\mu}=1.5$. The parameters are fixed to the base case values unless indicated otherwise. 
The droplet has initially a spherical shape, and we follow its time evolution under suddenly imposed constant shear (constant plate velocity). This process differs considerably depending on the droplet elasticity, which will be demonstrated by comparing the process two different Weissenberg numbers: $Wi=0.01$ (nearly viscoplastic droplet) and $Wi=0.5$ (finite elasticity). 
 
The time-evolution of the solid region for fully unyielded case at $Bi=0.8$ and $Wi=0.01$ is shown in Fig.~\ref{Solid0p25_0p01}. As can be seen in the figure, the near-viscoplastic droplet yields fully and fast (by $t=0.0189$), first remaining spherical. Eventually at $t>3$, the droplet has deformed by the shear flow and attained a nearly ellipsoidal shape. This result is intuitive, because the near-viscoplastic droplet behaves like a rigid solid and cannot deform unless fluidized. This leads to the observed steady states for near-viscoplastic droplets throughout the paper: they either contain solid regions at steady state but have not deformed from the spherical shape, or have fluidized completely and deformed to an elliptical shape. 
  
This can be compared to the yielding process of a droplet at the same Bingham number $Bi=0.8$ but at a larger Weissenberg number $Wi=0.5$, as depicted in Fig~\ref{Solid0p25_0p5}. In contrast to the near-viscoplastic case, the unyielding process is slow. The droplet first yields in the vicinity of the interface ($t=3.15$), then solid region starts to oscillate yielding and unyielding diagonally across the droplet (3.15$<t<$54.6), until it finally attains a steady state with two separate solid regions ($t=54.6$). The material yields fast at low $Wi$, whereas at larger $Wi$ the material is more elastic, leading to a slower yielding process and a larger solid region. Elasticity enables deformation at the solid state, which allows elastic and capillary stresses to redistribute inside the droplet and counteract the shear force in some regions. 
 
Solid (unyielded) regions in yield-stress fluids can either be stagnant (with zero velocity), such as fouling zones in porous media \citep{Chaparian_JNNFM_2019}, or they can move as solid plugs. In order to better understand the flow physics inside the droplet, where solid and fluidized regions may co-exist, we show the velocity vectors at equilibrium for the more elastic EVP droplet shown before ($Bi=0.8$ and $Wi=0.5$), along with a near-viscoplastic droplet in its completely unyielded state ($Bi=8$ and $Wi=0.01$) in Fig.~\ref{StreamBi}, right column. Clearly, in both cases the droplet experiences a tank-treading motion similarly to capsules and elastic particles in simple shear flow.  By comparing the velocity distributions with the unyielded zones shown in the left column, we observe moving unyielded zones for both regimes, i.e. the material keeps moving inside the droplet. For the near-viscoplastic droplet, the movement is a simple solid body rotation. For the more elastic EVP droplet, the material continuosly unyields and yields when passing through boundaries of solid zones, while the solid zones remain at same place in steady state. This shows that the solid regions of the droplet are \textit{pseudoplugs}, which have been observed in EVP flow in porous media \cite{DEVITA201810, Chaparian_JNNFM_2019}. We can also observe that the velocity vectors do not show any change in direction or character when entering or leaving the pseudoplug. This may seem unexpected, but is typical for pseudoplugs in yield-stress fluids. Pseudoplugs in cavity flows and cylinder flows of a Bingham fluid are shown in \cite{Saramito_complex}, together with streamlines of the velocity field. In all those examples, the velocity field seems largely unaffected by the presence of pseudoplugs.

It is worth noting that the tank-treading motion constrains the maximum deformation of a material element inside the droplet. This is fundamentally different from parallel shear flows, where infinite deformation is possible. The deformation constraint may partly account for the surprising yielding behaviours at increasing Weissenberg numbers, discussed in the next section.

%New figure 8: Unyielded regions Bi-Wi at equilibrium
\begin{figure}
\begin{center}
\begin{tabular}{c}
{\includegraphics[width=\textwidth]{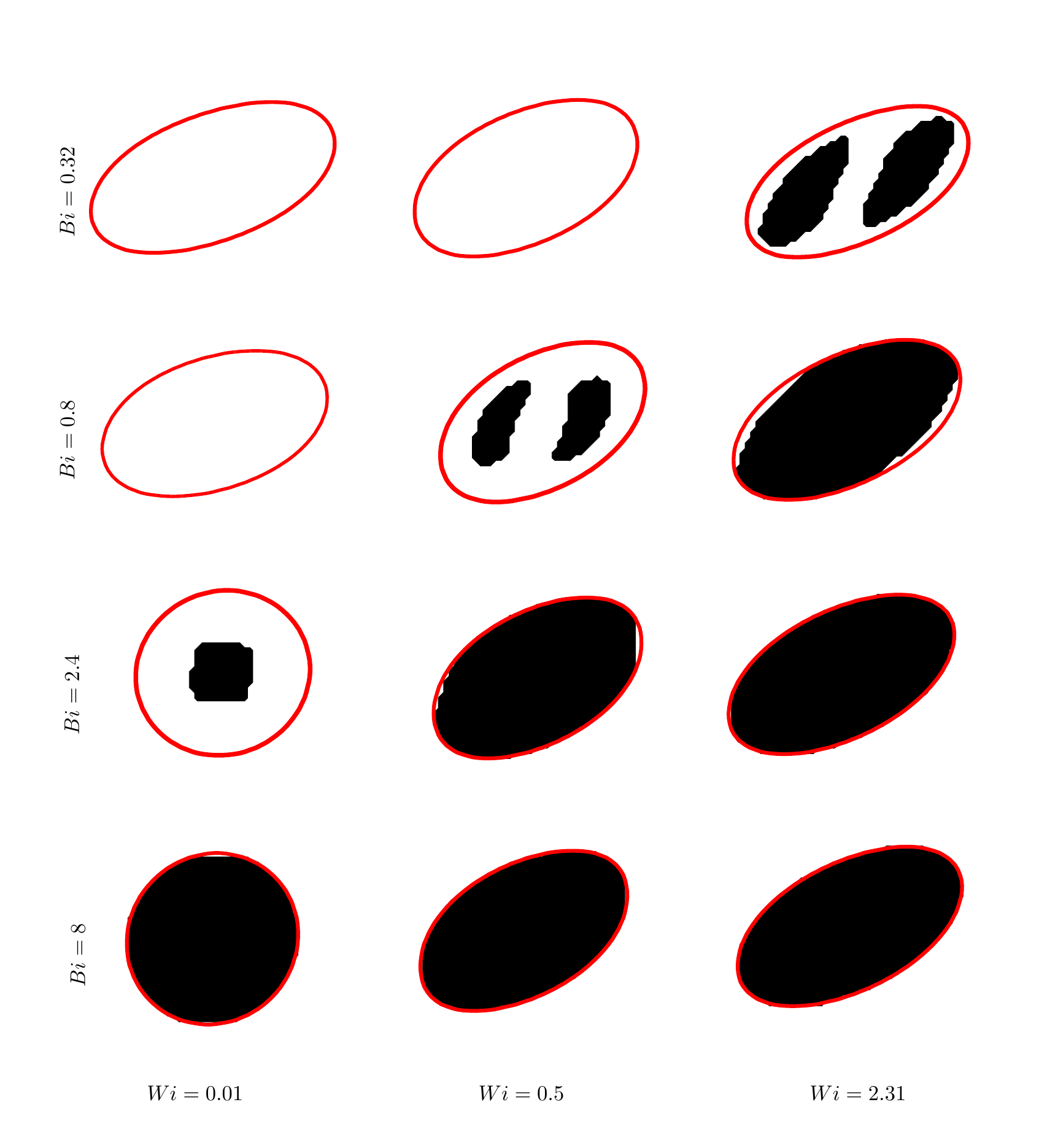}}
\end{tabular}
\end{center}
\caption{The effects of the Bingham and Weissenberg numbers on the unyielded regions inside the EVP droplet at equilibrium ($Bi$ and $Wi$ values given in the figure). The unyielded region is represented by the black colour at $\mathcal{F} \le 10^{-3}$, where $\mathcal{F}=1-(Bi/|\bm{\tau_{d}}|)$.}
\label{SolidBiWi}
\end{figure}

%New figure 9: Trace and normal stress difference, low vs. high Wi, at equilibrium
\begin{figure}
\begin{center}
\begin{tabular}{c}
{\includegraphics[width=\textwidth]{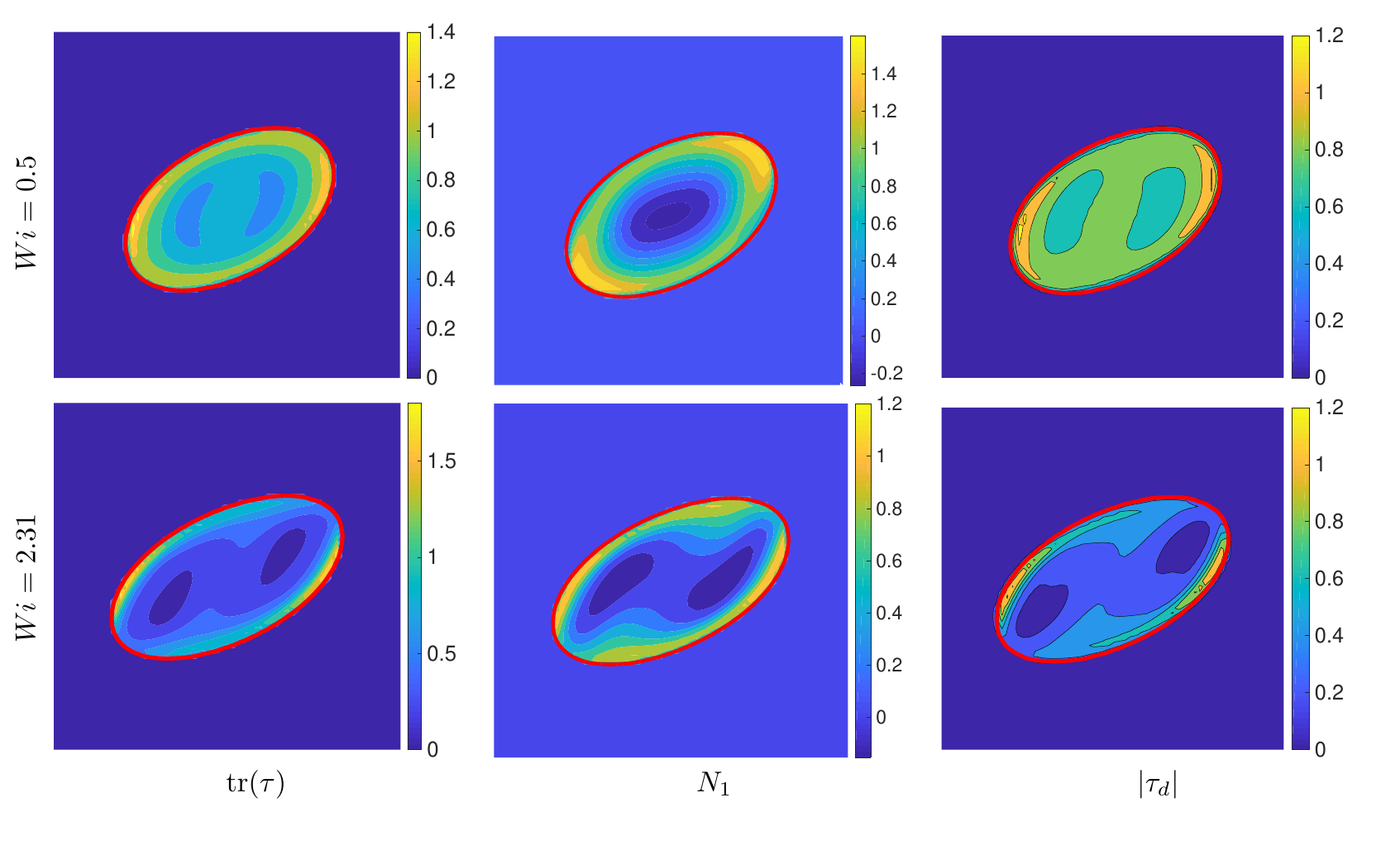}}
\end{tabular}
\end{center}
\caption{The effects of the Weissenberg number on the trace of the extra stress tensor $\rm{tr}(\bm{\tau})$ (left column), the first normal stress difference $N_1=\tau_{xx}-\tau_{yy}$ (middle column) and the deviatoric extra stress tensor $|\bm{\tau}_{d}|$ (right column) inside the EVP droplet at $Bi=0.8$. Weissenberg number increases from top to bottom (values of $Wi$ are given in the figure).}
\label{Tr_N}
\end{figure}

%New figure 11: Effect of Ca vs. effect of Wi on solid regions 
\begin{figure}
\begin{center}
\begin{tabular}{c}
{\includegraphics[width=\textwidth]{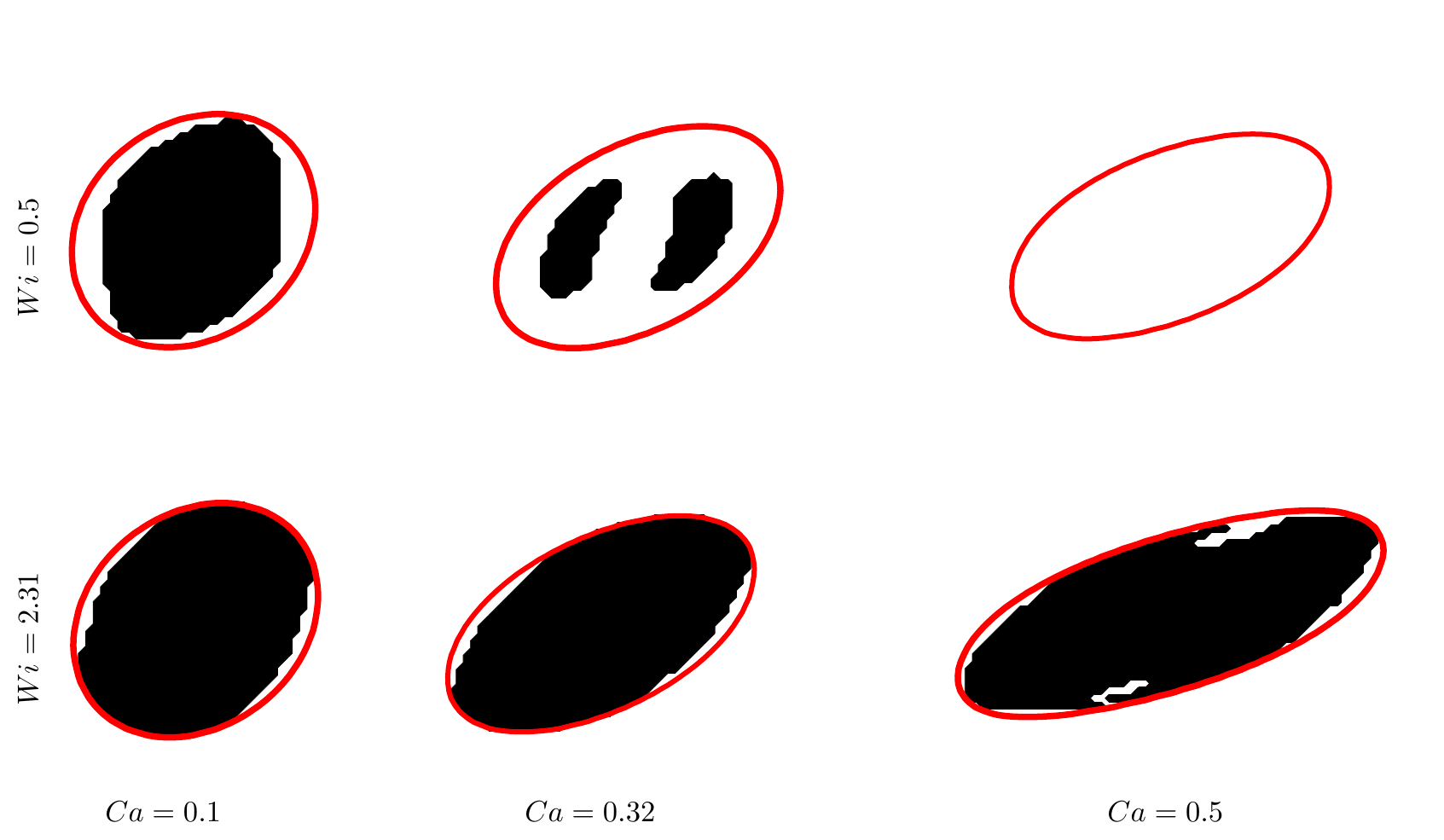}}
\end{tabular}
\end{center}
\caption{The effects of the capillary and Weissenberg numbers at $Bi=0.8$ on the yielded and unyielded regions inside the EVP droplet at the equilibrium. The unyielded region is represented by the black colour at value of $\mathcal{F}=10^{-3}$, where $\mathcal{F}=1-(Bi/|\bm{\tau_{d}}|)$.}
\label{Sol_Ca}
\end{figure}

%New figure 10: Regime map solid regions Wi-Bi
\begin{figure}
\begin{center}
\begin{tabular}{c}
{\includegraphics[width=0.65\textwidth]{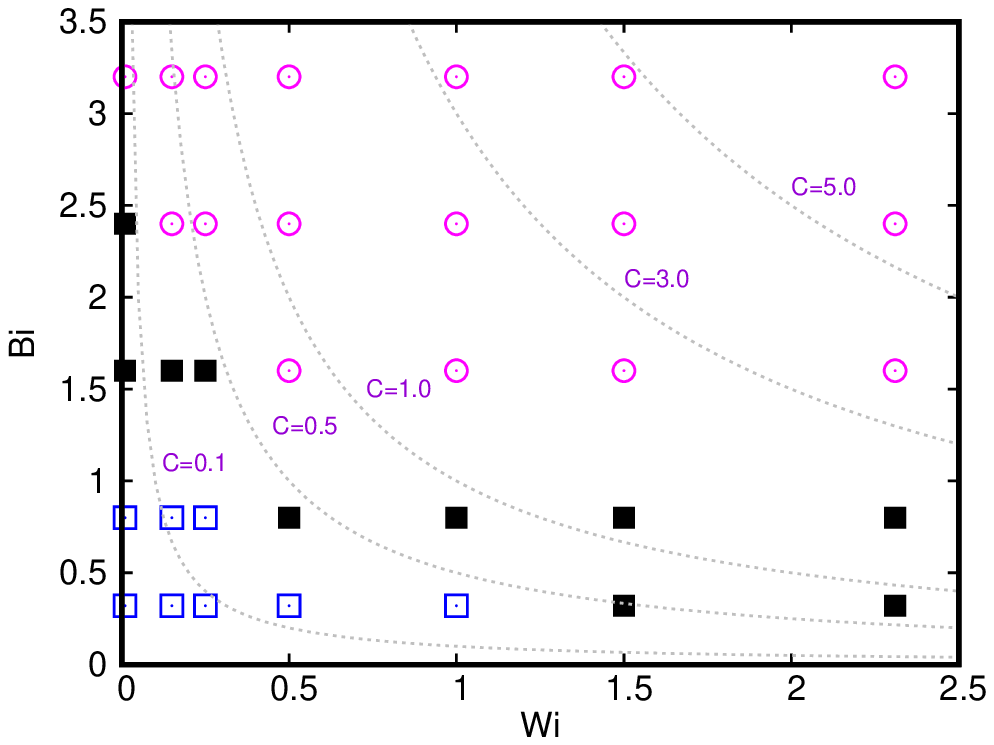}}
\end{tabular}
\end{center}
\caption{A regime map for the fully yielded ($\Box$), partly yielded ($\blacksquare$) and unyielded regimes ($\bigcirc$) of the EVP droplet with respect to $Wi$ and $Bi$. Dashed lines represents solution for constant material curve $Bi\times Wi=C$ at $C=0.1,0.5,1,3$ and 5.}
\label{MapSolid}
\end{figure}

\subsection{Effect of Weissenberg and capillary numbers}

The dynamics of an elastoviscoplastic drop is governed by a subtle interplay between various competing effects, and we move further to explore their respective roles. First, we look at how the distribution of fluid (yielded) and solid (unyielded) regions inside the droplet changes with the Bingham number and the Weissenberg number, all other parameters kept constant. Figure~\ref{SolidBiWi} shows the droplet interface at steady state, together with the unyielded regions, at twelve different parameter combinations of $Bi$ and $Wi$. It is worth remembering that the left column contains only near-viscoplastic droplets and that elasticity increases to the right in the figure.

The drop at the upper left corner has a combination of a low yield stress and small elasticity ($Bi=0.32$ and $Wi=0.01$), and is fully yielded. The Bingham number increases downwards in the figure, and the droplets in the lowermost row which have a high yield stress ($Bi=8$) are fully unyielded. Correspondingly, the droplets at an intermediate yield stress ($Bi=0.8$ and 2.4) contain a mix of yielded and unyielded regions. This development with Bingham number agrees with expectations: the droplet becomes more solid when the yield stress increases. Note that at intermediate and large yield stresses $Bi=2.4$ and 8 and low $Wi=0.01$ the droplet remains nearly spherical even for the case when it is partly yielded, cementing the qualitative difference between near-viscoplastic and EVP droplets in the previous section. It is more interesting to confirm that the solid regions also increase in volume with the Weissenberg number (going from left to right in the figure). Note that a similar trend has been observed for pseudoplugs in other complex geometries~\cite{DEVITA201810, Chaparian_JNNFM_2019}. The droplets at $Bi=0.32$ and $Bi=0.8$ are both fully yielded at $Wi=0.01$, but most of their volume is unyielded at $Wi=2.31$. Because our value of $\beta=0.68$ is higher than is typical for Carbopol, we have verified that this trend persists also for $\beta=0.05$. 

Moreover to explain how this happens, we now examine the stress fields at $Wi=0.5$ and compare them against those at $Wi=2.31$ at same Bingham number ($Bi=0.8$). Figure~\ref{Tr_N} compares the magnitudes inside the droplet of the trace of the extra stress tensor $\rm{tr}(\bm{\tau})$ (left column), first normal stress difference $N_1=\tau_{xx}-\tau_{yy}$ (middle column) and the deviatoric extra stress tensor $|\bm{\tau}_{d}|$ (right column). The maximum value of $\rm{tr}(\bm{\tau})$ increases slightly with $Wi$, while the maximum value of $\bm{\tau}_{d}$ is unchanged, and for $N_1$ decreases slightly. However, the distribution of those maximum values confines to thinner region along the interface. Since $Bi/|\bm{\tau}_{d}|$ determines the yield threshold, when $\bm{\tau}_{d}$ decreases, the unyielded zones increase. The distributions of the $\rm{tr}(\bm{\tau})$ represent degree of local elastic deformation, which also varies more inside the droplet when $Wi$ is increased. These figures show that although the maximum stress is nearly unchanged with elasticity, the elastic droplet re-distributes the stress more efficiently, leaving large central regions at lower stress and lower deformation. 

We also note that in the Saramito model, the elastic stress in the unyielded regime can be written as \cite{saramito2007new, saramito2009new} $\tau_e=(\mu_p/\lambda)\epsilon$, where $\epsilon$ is the local deformation. The elastic stress is inversely proportional to the elastic time scale $\lambda$ (\textit{i.e.} $Wi$), if the local deformation $\epsilon$ is constant. Hence, the overall stress is expected to decrease with $Wi$, if the droplet overall deformation remains roughly constant. In our case, the droplet deformation is mainly determined by capillary forces (except in near-plastic cases at high $Bi$), and therefore remains similar when $Wi$ increases. Hence, the elastic stresses are indeed expected to decrease overall with $Wi$, and the unyielded regions increase overall. As a secondary effect, it is also likely that capillary stresses at the droplet tips counteract better the external shear for the elastically deformable droplet; for the rigid near-viscoplastic droplet, the yield stress has to act as the main balancing force. To support this hypothesis, the unyielded regions are shown in Fig.~\ref{Sol_Ca} at different capillary numbers for droplets at two representative Weissenberg numbers ($Wi=0.5, 2.31$) and fixed $Bi=0.8$. The capillary number increases when going from left to right in the figure, and is varied in the range of  $0.1\le Ca\le0.5$. Indeed, increasing $Ca$ and resulting weakening of the capillary force results in an earlier yielding. A higher tip curvature is required to balance the external shear when capillary forces are weak. This implies larger deformation of the droplet, which results in higher elastic stresses than at low $Ca$, and therefore yielding. At $Wi=0.5$, the material is not elastic enough to sustain solid regions with increase in deformation with $Ca$, while at $Wi=2.31$, the droplet remains mostly solid. We also note that unyielded regions increase with $Wi$ at low $Bi$ for all $Ca$ considered in the current work.  

A complete map of the yielded and unyielded regimes is shown in Fig.~\ref{MapSolid} for $Wi$ and $Bi$ in the range of $0\le Wi \le 2.31$ and $0\le Bi \le 3.2$, respectively. Each marker in this figure corresponds to one simulation, where the final state of the droplet has been classified into one of three regimes: i) fully yielded (hollow square), ii) partly yielded (filled square), or iii) fully unyielded (circle). The partly yielded regime denotes a droplet containing yielded and unyielded regions simultaneously. For low and moderate Bingham numbers ($0<Bi<1.6$), the droplets are in the yielded or partly yielded regimes. For high Bingham numbers however ($Bi>2.4$), we find that the droplet is in the fully unyielded regime for all $Wi$ studied. For each value of $Wi$, a critical Bingham number $Bi_c$ can be defined as the lowest Bingham where the droplet enters the partly yielded regime. The figure shows that the critical Bingham number for partly yielded regime, $Bi_c$, decreases as a function of the Weissenberg number. This confirms that the trend observed in figure \ref{SolidBiWi} is universal; the unyielded region increases in volume and penetrates further within the bulk of the drop with $Wi$. Note that in an experiment, for a given material, an increase in the shear rate results in decreasing $Bi$, but an increasing $Wi$ and $Re$. It is possible to keep the capillary number constant experimentally by changing the droplet size. Moreover, in some experiments inertial effects are negligible. To mimic this in Fig.~\ref{MapSolid}, we also included constant material curves $Bi\times Wi =C$ at $C=0.1,0.5,1,3$ and 5, where different values of $C$ refers to various materials. Thus following each curve from left to right implies increasing shear rate for same material, while keeping capillary number constant. On the other hand, horizontal movement between two curves implies changing the material elasticity while keeping the same shear. All droplets yield when the shear increases sufficiently, but more elastic EVP droplets yield at higher levels of shear than less elastic or near-viscoplastic droplets.

\section{Comparison between EVP and purely viscoelastic droplets  \label{sec:comparison}}
While the previous section focused on the plastic and elastic effects and solid distributions of EVP droplets (information hard to find experimentally), the aim of this section is to shed light on the differences between EVP and viscoelastic droplets. The two quantities analysed here are relatively easy to measure and compare in possible future experiments: droplet deformation and orientation. 

%New figure 12: Drop deformation vs. Wi
\begin{figure}
\begin{center}
\begin{tabular}{c}
{\includegraphics[width=\textwidth]{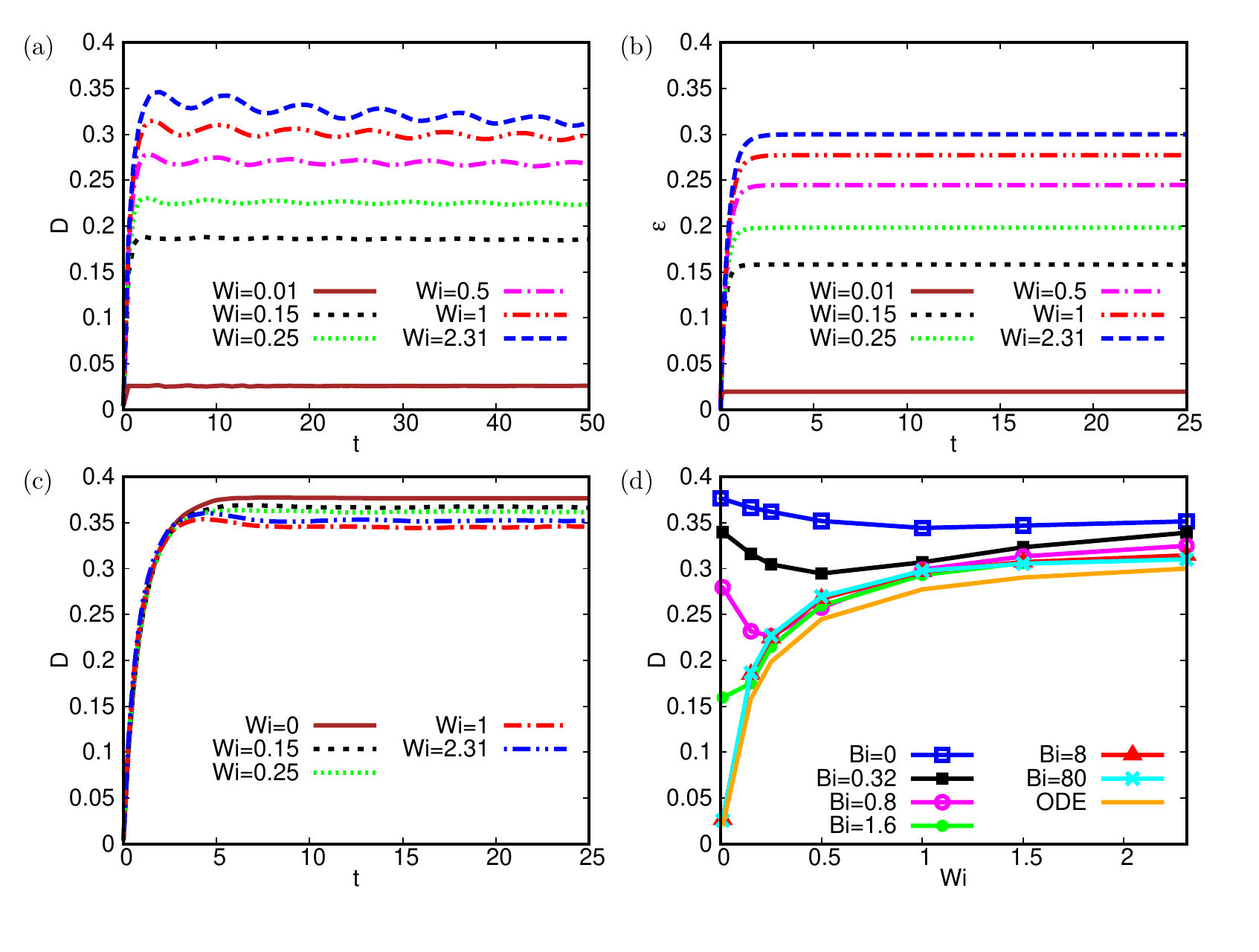}}
\end{tabular}
\end{center}
\caption{Effects of the Weissenberg and the Bingham numbers on the drop deformation. Top row shows transient solution for EVP droplet at $Bi=8$: (a) numerical solution and (b)  the ODE model prediction. Bottom row shows (c) the transient solution for VE droplet ($Bi=0$) and (d) long-time deformation $D$.}
\label{WiTrD}
\end{figure}

%New figure 14: Drop orientation vs. Wi
\begin{figure}
\begin{center}
\begin{tabular}{c}
{\includegraphics[width=\textwidth]{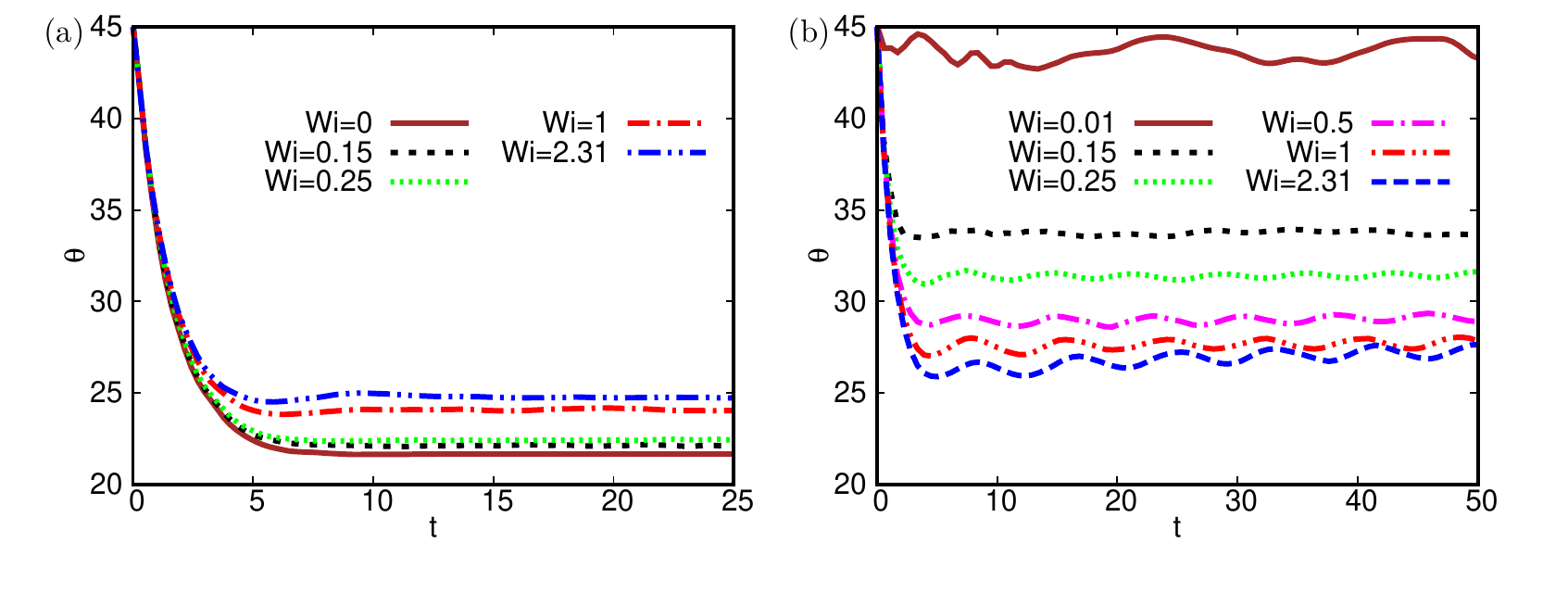}}
\end{tabular}
\end{center}
\caption{Effects of the Weissenberg number on the orientation angle $\theta$. Left (a) and right (b) represent $Bi=0$ and $Bi=8$ cases, respectively. The temporal evolution of $\theta$ is shown for values of $Wi$ in the range $0\le Wi \le 2.31$. }
\label{WiTrAng}
\end{figure}

\subsection{Effect of the Weissenberg number \label{sec:deformationWi}}

Figures~\ref{WiTrD} (a)-(d) compare the deformation of a purely viscoelastic ($Bi=0$) and an EVP droplets. In (a), the time evolution of the deformation parameter $D$ is shown for different values of $Wi$ in the range $0 \le Wi \le 2.3$, for a fully unyielded EVP droplet at $Bi=8$. This result shows that the droplet deformation oscillates periodically in time, with a nearly constant amplitude. Indeed, we performed the simulations over a long time period (t=100) and did not observe changes in frequency or amplitude for $Wi>0.15$. However, the oscillations decrease in amplitude when the Weissenberg number is lowered. At low to intermediate $Wi$ ($Wi \le 0.15$), the periodic oscillations and the drop deformation are probably suppressed by the large stiffness of the material (a high elastic spring constant $k_e=(1-\beta)/Wi$). Correspondingly, at high $Wi$, the elasticity $k_e$ decreases resulting in a larger mean deformation and an enhanced oscillation. 

This qualitative development is sharply contrasted by the VE droplet in Fig.~\ref{WiTrD}(c). The deformation of the VE droplet in time exhibits a slight overshoot before saturating to a steady state value. The overshoot appears because a finite time is required to develop inhibitive viscoelastic stresses, and is therefore more pronounced for higher $Wi$ (when the characteristic time scale is longer). 
The steady-state value of the drop deformation changes non-monotonically with $Wi$, which is again in agreement with the numerical results of \citet{Aggarwal_JFM_2007}.  When considering the saturated mean value of the drop deformation, in agreement with previous studies we observe that increasing $Wi$ \textit{decreases} the deformation of the VE droplet, in contrast to the EVP droplet for which the deformation \textit{increases} with viscoelasticity. 

Figure~\ref{WiTrD}(b) shows the prediction of the ODE model for an EVP droplet presented in this paper (Appendix~\ref{sec:ODE}), with a surface tension spring and an elastic spring in parallel with a viscous damper. We would like to emphasize that the model describes a fully unyielded droplet, which is the case also in the numerical simulation in (a). The model captures the development of the mean deformation with $Wi$ very well, indicating that the droplet at $Bi=8$ indeed behaves like a viscoelastic solid. However, the ODE model cannot predict the time-periodic oscillations. We calculated the natural frequency of the corresponding spring-damper system, and found it to be almost an order of magnitude higher than the frequency observed in our simulations that was close to the flow shear time scale ($t=1$). However, at $Re=0.05$, the shear flow itself is very stable. Hence, we propose that the oscillations are a result of a fluid-structure interaction, rather than determined by either phase alone. It is worth mentioning that similar periodic oscillations were observed for viscoelastic droplets which were much more viscous than the outer fluid \citep{Mukherjee_JNNFM_2009}. Here, although the viscosity ratio is low ($k_\mu=1.5$), the droplet behaves like a very viscous one because it is fully solid.   

Finally, the development of the saturated (mean) values of the drop deformation with $Wi$ is quantified in Fig.~\ref{WiTrD}(d) for a wide range of Bingham numbers ($Bi=0-80$). This figure shows how the dependence on $Wi$ changes continuously from non-monotoneous but mainly decreasing at $Bi=0$, to monotoneous and highly increasing at $Bi \ge 1.6$. One can observe that the viscoelastic effects are more pronounced in the unyielded regime. The long-time drop deformation at high $Bi$ increases monotonically with the Weissenberg number and eventually reaches a plateau. The ODE prediction (orange line) is also included. Interestingly, the numerical simulation results at high $Bi$ are close to the predictions of the ODE model, as might be expected as the model was derived for the fully unyielded regime. We cannot expect them to fully converge to each other, because of the extreme simplicity of the model. 

In Fig.~\ref{WiTrAng}, the time development of the orientation angle $\theta$ is shown for various values of $Wi$ in the range $0\le Wi \le 2.31$. The time evolution of the orientation angle is similar to the drop deformation, i.e., the EVP droplet approaches the steady state orientation significantly slower than the in viscoelastic droplet, due to overshoots accompanied with oscillations in the EVP case. Next, we will attempt to interpret the steady state values. It is known that the deformation of a Newtonian droplet increases with $Ca$, and that the droplet therefore aligns itself closer to the flow direction (smaller $\theta$). The viscoelastic droplet (Fig.~\ref{WiTrAng}(a)) decreases its deformation with $Wi$, and accordingly, $\theta$ increases (further away from the flow direction). This observation is in agreement with \citet{Aggarwal_JFM_2007}. For the EVP droplet on the other hand (Fig.~\ref{WiTrAng}(b)), the trends are reversed: the drop alignment with the flow increases with $Wi$. The change in the EVP droplet orientation leads to changes in the external flow field, and increases the flow-induced viscous stretching of the EVP droplet, as further discussed in next section. 

%New figure 13: Stress oscillation in time
%\begin{figure}
%\begin{center}
%\begin{tabular}{ccc}
%{\includegraphics[width=0.3\textwidth]{figs/N1_2p31_9000000.eps}} &
%{\includegraphics[width=0.3\textwidth]{figs/N1_2p31_9100000.eps}} &
%{\includegraphics[width=0.3\textwidth]{figs/N1_2p31_9200000.eps}} \\
%$t=94.5$ & $t=95.55$ & $t=96.6$ \\
%{\includegraphics[width=0.3\textwidth]{figs/N1_2p31_9300000.eps}} &
%{\includegraphics[width=0.3\textwidth]{figs/N1_2p31_9400000.eps}} &
%{\includegraphics[width=0.3\textwidth]{figs/N1_2p31_9500000.eps}}\\
%$t=97.65$ & $t=98.7$ & $t=99.75$
%\end{tabular}
%\end{center}
%\caption{Time evolution of the elastoviscoplastic droplet ($Bi=1.6$, $Wi=2.31$) in the shear plane ($x-y$ plane at $z=L_z/2$). The contours represent the first normal stress difference $N_1=\tau_{xx}-\tau_{yy}$.}
%\label{N1Tr}
%\end{figure}

% New figure 15: Stresses at the edge vs. Wi
\begin{figure}
\begin{center}
\begin{tabular}{c}
{\includegraphics[width=\textwidth]{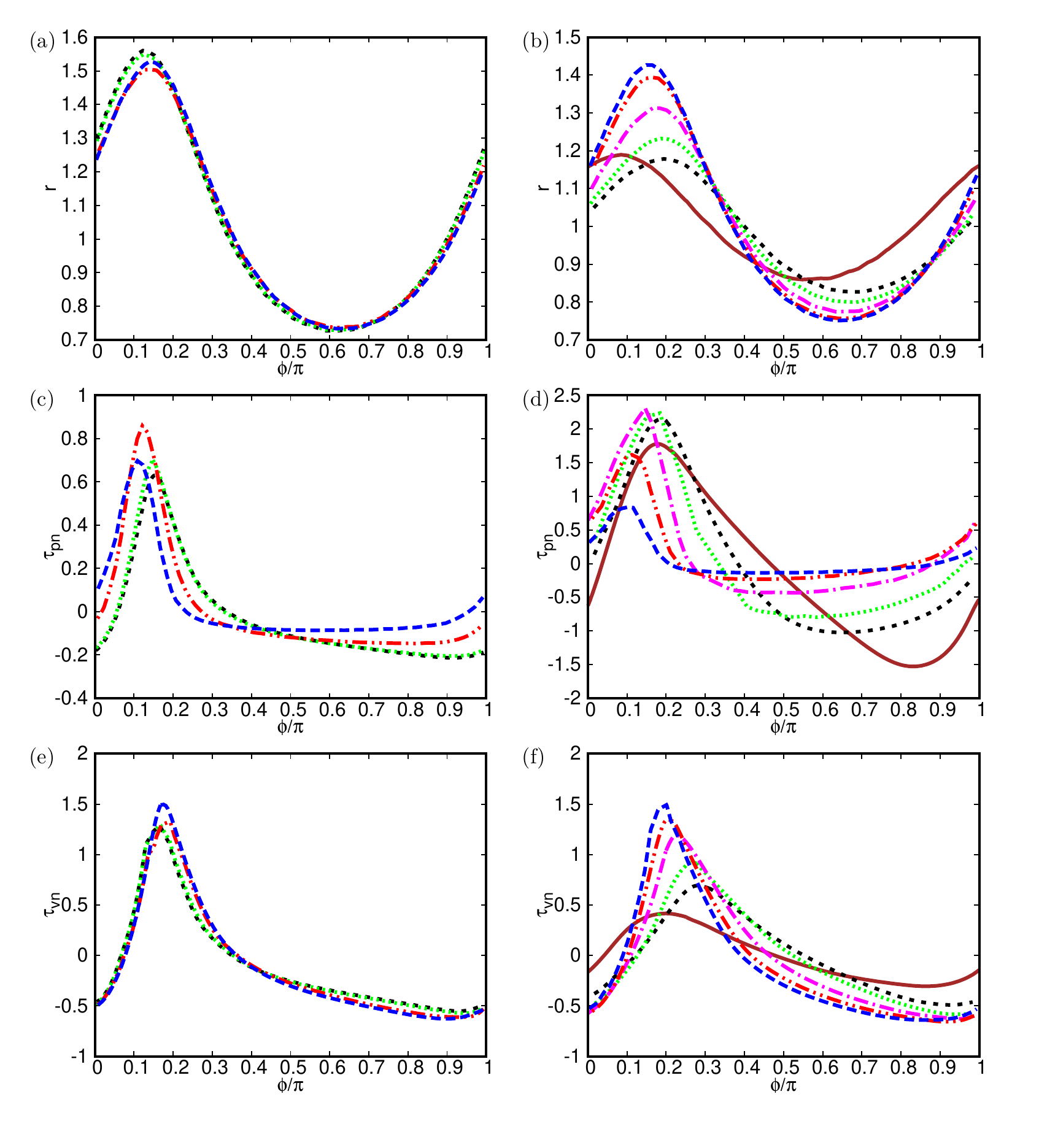}}
\end{tabular}
\end{center}
\caption{The drop interface in polar coordinates for (a) $Bi=0$, and (b) $Bi=1.6$. Stresses along the inner edge of the interface for different $Wi$: (c) EVP normal stress ${\rm \tau_{pn}}=\textbf{n} \cdot \pmb{\tau} \cdot \textbf{n}$ for $Bi=0$, (d) EVP normal stress for $Bi=1.6$, (e) Viscous normal stress for $Bi=0$, (f) Viscous normal stress for $Bi=1.6$. All stresses are shown in the $z=L_z/2$ plane versus the angular position $\phi$, at different Weissenberg numbers in the range $0\le Wi \le 2.31$. 
The zero angular position coincides with the $x$-axis. The brown (\protect\tikz[baseline]{\protect\draw[line width=0.5mm][color=brown] (0,.6ex)--++(0.5,0) ;}), black (\protect\tikz[baseline]{\protect\draw[line width=0.5mm,dash pattern=on 1mm off 1mm] (0,.6ex)--++(0.5,0);}), green  (\protect\tikz[baseline]{\protect\draw[line width=0.5mm,dotted] [color=green](0,.6ex)--++(0.5,0);}), magenta  (\protect\tikz[baseline]{\protect\draw[line width=0.5mm,dashdotted][color=magenta] (0,.6ex)--++(0.5,0);}), red (\protect\tikz[baseline]{\protect\draw[line width=0.5mm,dashdotdotted][color=red] (0,.6ex)--++(0.5,0);})  and blue (\protect\tikz[baseline]{\protect\draw[line width=0.5mm, dash pattern=on 2mm off 1mm][color=blue](0,.6ex)--++(0.5,0);}) colors are used for $Wi=$ 0.01, 0.15, 0.25, 0.5, 1 and 2.31, respectively.
}
\label{Bi0Circ}
\end{figure} 

\subsubsection{Physical explanation of the effect of Weissenberg by normal stress differences \label{sec:emulsions}}

%The time evolution of the first normal stress difference $N_1=\tau_{xx}-\tau_{yy}$ inside the elastoviscoplastic drop is illustrated in Fig.~\ref{N1Tr} in the shear plane ($x$-$y$ plane at $z=L_z/2$) for the case with $Bi=1.6$ and $Wi=2.31$. It can be seen that the maximum values of $N_1$ are distributed along the interface, except for the regions near the leading and trailing edge. The stress gradients are observed to fluctuate in the middle part of the drop. We find that the stress evolution is unsteady and periodic in time at this Bingham number, especially with increasing viscoelasticity. This phenomenon leads to oscillations of the drop deformation in time, considered in Section~\ref{sec:deformationWi}. %It is worth noting that \citet{fdv1} observed time-dependent oscillations in the EVP flow through porous media.   

We now consider the distribution of stresses along the droplet interface, following the arguments of \citet{Yue_JFM_2005} and \citet{Aggarwal_JFM_2007} who studied the same for viscoelastic droplets. The stress fields (Fig.~\ref{Bi0Circ}(c)-(f)) and the drop interface (Fig.~\ref{Bi0Circ}(a),(b)) are shown in the central plane ($z=L_z/2$) as functions of the angle $\phi$ measured from the streamwise direction. By analysing these distributions, we seek to understand why the trends for VE and EVP droplets are different. 

The distribution of the EVP normal stress (\textit{i.e.} ${\rm \tau_{pn}}=\textbf{n} \cdot \pmb{\tau} \cdot \textbf{n}$, where $\textbf{n}$ is the outgoing normal of the interface) is shown for $Bi=0$ (Fig.~\ref{Bi0Circ} c) and $Bi=1.6$ (Fig.~\ref{Bi0Circ} d). At both Bingham numbers they share some features. The maximum of the EVP normal stress lies in both cases in the vicinity of the drop tips while its minimum lies around the equator. Furthermore, ${\rm \tau_{pn}}$ is tensile near the drop poles, thus pulling the interface inward and reducing the long $L$-axis, whereas near the equator region it is compressive and pushes the interface outward, increasing the short $B$-axis (figure \ref{Bi0Circ}c,d). Hence, the stresses at the tips and at the equator act mainly to reduce the droplet deformation.

Considering now the purely viscoelastic droplet ($Bi=0$), the tensile stress at the tip shows a non-monotonic trend. In increasing the Weissenberg number from $Wi=0.15$ to $Wi=1$, the maximum value of the tensile stress increases. However, on further increasing it to $Wi=2.31$, the peak value slightly decreases. Because of this action, an increase in $Wi$ results in a non-monotonic trend of the drop deformation, i.e. the deformation first decreases and then slightly increases.

For the elastoviscoplastic droplet ($Bi=1.6$), the tensile stress undergoes overall a much larger variation with increasing $Wi$. The variation at the tips is still non-monotonic, but the compressive stress around the equator affecting the $B$-axis decreases monotoneously and significantly with $Wi$. Because of this, the deformation of the EVP droplet increases monotoneously with $Wi$. 

Finally, the viscous normal stress is shown in Fig.~\ref{Bi0Circ}(e,f) (i.e. ${\rm \tau_{vn}}=\textbf{n} \cdot \mu_{s}(\nabla\textbf{u}+\nabla\textbf{u}^T) \cdot \textbf{n}$).  In general, a higher magnitude of normal viscous stress at the drop top increases $L$, while at the equatorial region it decreases $B$, both resulting in larger deformation. For the EVP droplet, it is found that the magnitude of the ${\rm \tau_{vn}}$ increases both at the drop-tip and the equator with Weissenberg, especially for the $Bi=1.6$ case. The drop aligns much more with the streamwise direction with $Wi$, thus, flow modifications induced by the EVP stresses result in increased viscous stretching. The VE droplet on the other hand ($Bi=0$) just slightly changes its inclination angle to be more aligned with the extension axis of the external flow.

Summarizing what the stress distributions tell us, we observe a non-monotonic deformation due to stress distribution at the drop tips and a simultaneous increase in deformation as a result of decreasing stress at the equator. The relative magnitudes of these two phenomena cause the non-monotonic trend for $Bi=0$ case and a monotonic increase for $Bi=1.6$.  

\subsection{Effect of capillary number and solvent viscosity ratio \label{sec:deformationCa}}

\begin{figure}
\begin{center}
\begin{tabular}{c}
{\includegraphics[width=\textwidth]{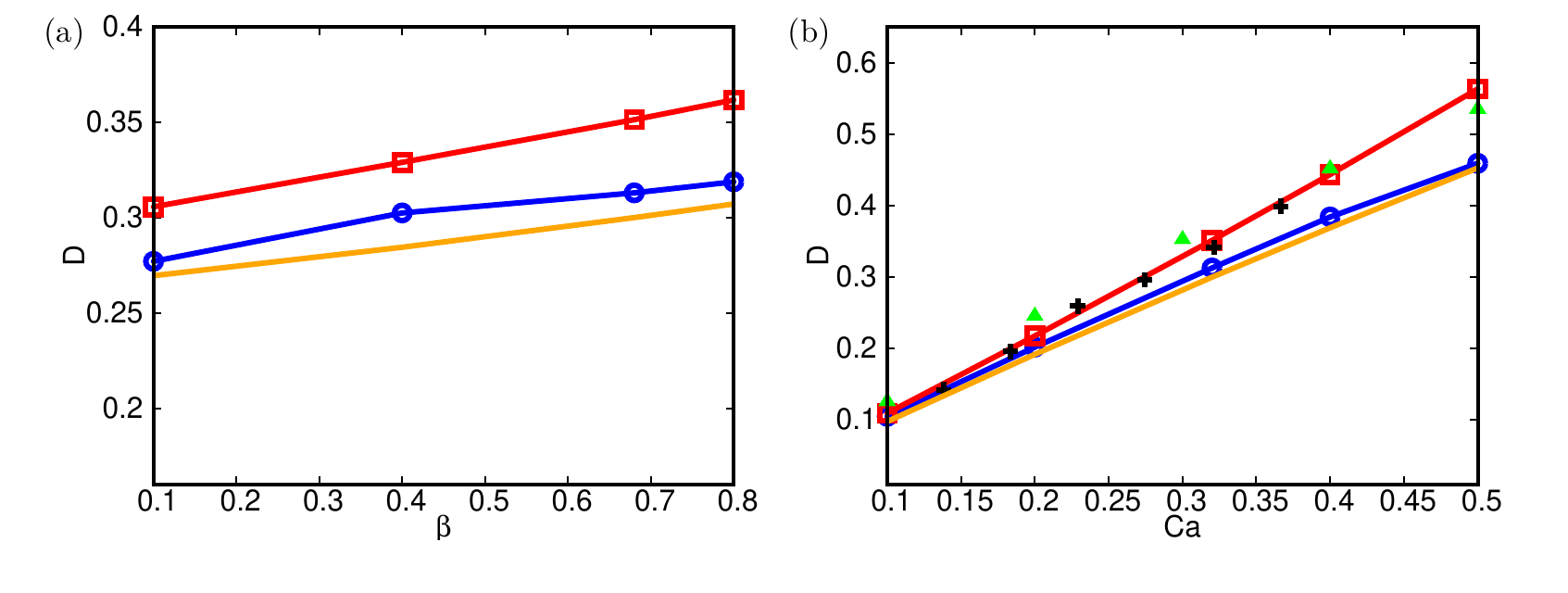}}
\end{tabular}
\end{center}
\caption{Long-time deformation parameter $D$ (a) {\it vs} $\beta$ in the range $0.1\le \beta \le 0.8$ and (b) {\it vs} $Ca$ for various values of $Ca$ in the range $0.1\le Ca \le 0.5$. Orange line: ODE model for EVP fluid; red line with $\Box$ symbols and blue line with $\circ$ symbols are results for $Bi=0$ and $Bi=1.6$, respectively; black $+$ symbol: experimental data from Ref. \cite{Verhulst_JNNFM_2009}; green $\bigtriangleup$ symbol: numerical results from Ref. \cite{Izbassarov_IJNMF_2018} (The Weissenberg number is $Wi=2.31$.)}
\label{BetaStD}
\end{figure}

Next, we examine how the solvent-to-polymeric viscosity ratio $\beta$ affects the dynamics of the droplet. In Fig.~\ref{BetaStD}, the long-time saturated mean values of the drop deformation are depicted for various values of $\beta$ in the range  $0.1\le \beta \le 0.8$, while keeping the droplet-to-ambient viscosity ratio $k_{\mu}$ constant. For the viscoelastic droplet in Fig.~\ref{BetaStD}(a), when $\beta$ is increased, the long-time value of $D$ increases. This can be explained by that $\beta$ modifies the effective Weissenberg number $Wi'=Wi(1-\beta)$, following \citet{Izbassarov_POF_2016, Aggarwal_JFM_2008}; increasing $\beta$ has a similar physical effect for a viscoelastic droplet as decreasing $Wi$. Elasticity inhibits deformation for the viscoelastic droplet, hence decreasing $Wi'$ increases its deformation.

Also for the EVP droplet with $Bi=1.6$ (Fig.~\ref{BetaStD}(a)), on increasing $\beta$, the long-time deformation increases. By the earlier argument this seems counterintuitive, because $Wi'$ decreases here thus would lead to decrease in deformation for the EVP droplet. However, it should be observed that $Wi'$ is based on the first normal stress difference $N_1$ for the viscoelastic droplet \citep{Aggarwal_JFM_2008}, and in Section~\ref{sec:emulsions} we could observe that the EVP droplet normal stress behaves differently. We can instead employ a simple physical argument based on the mechanistic system that describes the droplet in its unyielded state (Appendix \ref{sec:ODE}). The nondimensional material elasticity parameter $k_e=(1-\beta)/Wi$ decreases with an increase in $\beta$. Therefore for the EVP droplet, increasing $\beta$ has a similar effect as increasing $Wi$, resulting in a increase of long-time deformation, differently from the viscoelastic droplet. The effective Weissenberg number for the fully unyielded EVP droplet then becomes $Wi'=Wi/(1-\beta)$. The proposed ODE model for the EVP droplet is also able to capture the increase in deformation (Fig.~\ref{BetaStD}a). Summarizing these results, an increase in $\beta$ increases the deformation for both viscoelastic and elastoviscoplastic droplets.

Finally, the effect of surface tension was investigated by varying $Ca$ in the range of $0.1\le Ca \le 0.5$. The long-time values (Fig.~\ref{BetaStD}(b)) of $D$ are shown. In Fig.~\ref{BetaStD}(b), the long-time deformation is compared with experimental data for a Boger fluid droplet in a Newtonian matrix \citep{Verhulst_JNNFM_2009}, and numerical results for an elastic particle in a Newtonian fluid system \citep{Izbassarov_IJNMF_2018}. The viscoelastic droplet agrees very closely with the other two, indicating that the surface tension has the same effect on all similar viscoelastic systems (an elastic particle could be considered as a viscoelastic material with an infinite relaxation time, hence very high $Wi$). It is worth noting that the long-time deformation of the EVP droplet becomes independent of Weissenberg number when $Wi$ is sufficiently high (Fig.~\ref{WiTrD}), which may explain the good match with the elastic particle. The deformation of the EVP droplet increases similarly with $Ca$, but the slope is slightly smaller than for the viscoelastic droplet, and again, the effect is well reproduced by the proposed ODE model.  

\section{Conclusions \label{sec:concl}}

A non-Newtonian elastoviscoplastic (EVP) droplet surrounded by a Newtonian simple shear flow was studied by three-dimensional direct numerical simulations. An elastoviscoplastic droplet exhibits simultaneously viscoelastic and plastic behaviours, and we studied its time-evolution under constant shear, and its steady state. First, we performed a numerical experiment with a droplet of Carbopol under increasing shear rates. The physics of the yielding process was then analysed by varying the nondimensional parameters of the problem. The yielding process in time for a near-viscoplastic EVP droplet ($Wi=0.01$) was much faster and qualitatively different from more elastic EVP droplets. Also, the unyielded regions (solid plugs and pseudoplugs) within the droplet increased in size with increasing Weissenberg number. This can be explained by that the elastic droplet can deform and redistribute its stresses more efficiently in the unyielded state, leaving large regions under low stress, and allowing capillary forces to act at the drop tips. As unyielded regions are larger at higher $Wi$, this leads to a somewhat surprising conclusion: plastic effects are enhanced by droplet elasticity. A two-dimensional regime map was constructed as a function of Weissenberg and Bingham numbers, classifying each case into one of three regimes, depending on whether the droplet is i) fully yielded, ii) partly unyielded of iii) fully unyielded. In summary, our simulations predicted that the volume of the unyielded region inside the droplet increases with the Bingham number and the Weissenberg number, while it decreases with the capillary number at low Weissenberg and Bingham numbers. 

Furthermore, we compared droplet deformation between viscoelastic and EVP droplets, a quantity which would be easy to measure in experiments. We found that the EVP droplet behaves in many ways differently from a viscoelastic droplet, but with a continuous transition between the two behaviours as a function of the Bingham number. Firstly, the deformation and stresses of the EVP droplet oscillated periodically in time. Such oscillations are not seen for viscoelastic droplets, but have been observed in systems with a high viscosity contrast. Secondly, the EVP droplet deformed more with an increasing Weissenberg number, in contrast with the viscoelastic droplet. The drop inclination angle also had opposite trends for VE and EVP droplets (not shown for brevity). However, both droplets deformed more with an increasing capillary number, and with a decreased polymeric viscosity (or more precisely, an increasing solvent-to-total viscosity ratio). While the dynamics of the viscoelastic droplet is governed by its first normal stress difference, the dynamics of the EVP droplet in its fully unyielded regime is mainly governed by its elastic spring constant.  Finally, we proposed a simple ordinary differential equation, which captures well the observed trends of the EVP droplet parameter dependence in the fully unyielded regime. However, the model was not able to capture the time-periodic oscillations observed in our numerical simulations. 

In summary, while any droplet of a yield-stress fluid will yield at high enough shear (low enough Bingham number), we found that also other parameters have a significant qualitative influence on its yielding behaviour and deformation. In particular, droplet elasticity and plasticity both need to be accounted for. 

In the future, we would welcome experimental studies of EVP droplets in shear flows, to confirm to which extent the qualitative differences between VE and EVP droplets can be observed in experiments. It would also be interesting to perform studies of EVP droplets with further improved models, including other physical effects that complex fluids often represent, such as shear-thinning \citep{saramito2009new}, kinematic hardening \citep{dimitriou2019} or even thixotropy \citep{dimitriou2014comprehensive}.

\section{Acknowledgments}
We acknowledge financial support by the Swedish Research Council through grants No. VR2013-5789 and No. VR2017-4809. 
The authors acknowledge computer time provided by SNIC (Swedish National Infrastructure for Computing). 
This project has received funding from the European Research Council (ERC) under the European Union's Horizon 2020 research and innovation programme (grant agreement No 852529).
%
%
%
%\appendix
%\section{}

\appendix
\begin{figure}
\begin{center}
\begin{tabular}{c}
{\includegraphics[width=\textwidth]{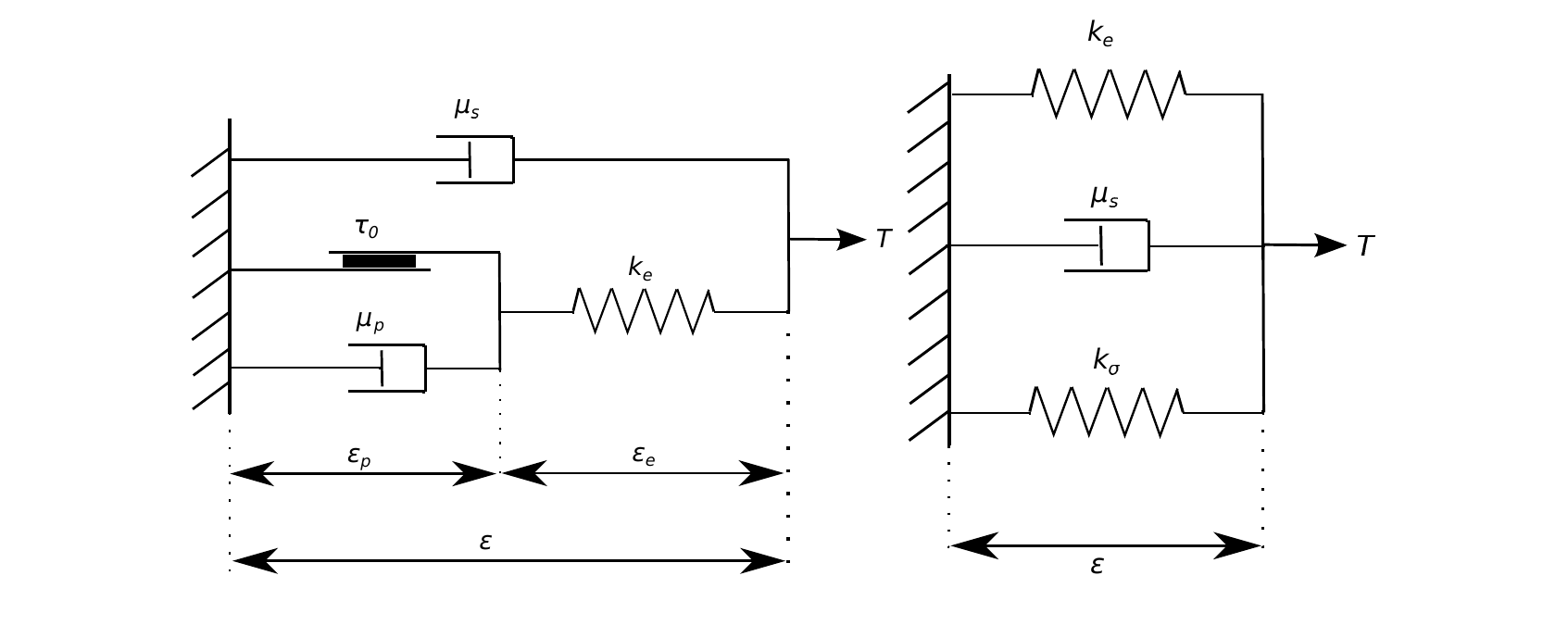}}
\end{tabular}
\end{center}
\caption{Sketch of the mechanical model for EVP fluid: (left) Saramito model for single phase fluid; (right) proposed model for an EVP droplet in its fully unyielded regime.}
\label{EVPmodel}
\end{figure}

\section{A phenomenological model for an EVP drop \label{sec:ODE}}

\citet{saramito2007new} proposed a mechanical system representing the physical behaviour of a single phase EVP fluid as illustrated in Fig.~\ref{EVPmodel}(a). When the local stress is less than the yield stress $\tau_0^*$, one obtains a recoverable Kelvin-Voigt system consisting of an elastic spring ($k_e^*=\mu_{p,2}^*/\lambda^*$)  and a viscous damper ($\mu_{s,2}^*$) in parallel. For stresses larger than the $\tau_0^*$, the friction element breaks and the system reduces to the Oldroyd-B viscoelastic model. The total strain rate $\dot{\varepsilon}^*$ is shared between the plastic contribution $\dot{\varepsilon_p}^*$  and  the elastic contribution $\dot{\varepsilon_e}^*$. The proposed model for an EVP droplet in the fully unyielded regime is illustrated in Fig.~\ref{EVPmodel}(b), \textit{i.e.} the stress is assumed to be less than the yield stress everywhere\footnote{It is mathematically possible to combine the two analytical models to allow partly yielded droplets. However, in order to provide physical results in this regime, the model would need to include an analytical a-priori estimate of the extent of yielded and unyielded regions as functions of different parameters.}. The surface tension at the deformable drop interface enters as a linear spring with a spring constant ($k_\Gamma^*$), in parallel with the Saramito elastic spring and the viscous damper. 
The total stress is obtained as the sum of the stresses of each element, in dimensional form:

\begin{equation}
\mu_{s,2}^* \dot{\varepsilon}^* + k_\Gamma^* \varepsilon^* + k_e^* \varepsilon^* = T^*.
\label{myODE}
\end{equation}
  
\citet{Aggarwal_JFM_2007} developed a first-order ordinary differential equation to describe the deformation of a viscoelastic drop in a steady shear flow, while here, a similar model is presented to explain the EVP drop deformation in the same flow. Eq.~\ref{myODE} can be rewritten in terms of forces by multiplying it with the area ${R^*}^{2}$ as: \begin{equation}
\underbrace{\mu_{s,2}^* {R^*}^{2} \dot{\varepsilon}^*}_{F_{viscous}} + \underbrace{k_\Gamma^* {R^*}^{2} \varepsilon^*}_{F_{interfacial}} 
+ \underbrace{k_e^* {R^*}^2 \varepsilon^*}_{F_{EVP}} = \underbrace{{\mu}_1^* {R^*}^{2} \dot{\gamma}^*}_{F_{total}}  
\label{ODE_eq},
\end{equation}
\noindent where the spring constant $k_\Gamma^*$ is defined as $k_\Gamma^*=\Gamma^*/R^*$. Eq.~\ref{ODE_eq} represents a force balance between the imposed shear force from the outer flow (right-hand side), and the viscous, interfacial and EVP forces of the droplet (left-hand side). On the right hand side, $F_{total}={\mu}_1^* {R^*}^2  \dot{\gamma}^*$ term is the viscous stretching of the droplet by the outer shear flow. On the left hand side, $F_{viscous}=\mu_{s,2}^* {R^*}^2 \dot{\varepsilon}^*$ is the viscous damping force inside the droplet, $F_{interfacial}=k_\Gamma^* {R^*}^2 \varepsilon^*$ is the force due to the surface tension, and the last term on the left $F_{EVP}=k_e^* {R^*}^2 \varepsilon^*$ is the contribution of elastoviscoplastic stresses. 
In order to solve the ODE model, an initial condition is required, and an initially spherical drop $\varepsilon^*(0)=0$ is assumed here (as in our numerical simulations). Eq.~\ref{ODE_eq} can be non-dimensionalized with the length scale $R^*$ and with the time scale 
$\dot{\gamma^*}^{-1}$, thus obtaining

\begin{equation}
\frac{{\rm d} \varepsilon}{{\rm d} t} +\frac{1}{\beta} \left \{\frac{1}{k_{\mu}}\frac{1}{Ca}+\frac{1-\beta}{Wi} \right \}
\varepsilon = \frac{1}{\beta}\frac{1}{k_{\mu}},  
\label{ODE_SRM}
\end{equation}

\noindent where $Ca=\mu_1^* R^* \dot{\gamma}^*/\Gamma^*$, $Wi=\lambda^* \dot{\gamma}^*$, $\beta=\mu_{s,2}^*/\mu_2^*$, $k_{\mu}=\mu_2^*/\mu_1^*$ and $\varepsilon$ are the model capillary number, the model Weissenberg number, the ratio of the solvent viscosity to the total viscosity, the ratio of droplet to bulk viscosity, and the drop deformation, respectively. 

An analytical solution can be obtained from (\ref{ODE_SRM}), i.e.

\begin{equation}
\varepsilon(t)=\frac{1}{\frac{(1-\beta)k_{\mu}}{Wi}+\frac{1}{Ca}} \left \{1-{\rm e}^{- \left (\frac{1-\beta}{Wi}+\frac{1}{Ca}\frac{1}{k_{\mu}}
\right ) \frac{t}{\beta}} \right \}.
\end{equation}
and the steady-state solution is
\begin{equation}
\varepsilon(\infty)=\frac{Wi Ca}{(1-\beta)k_{\mu}Ca+Wi}.
\label{ST_ST}
\end{equation}

%Note that as it was mentioned by Ref. \cite{Aggarwal_JFM_2007} the proposed ODE models are only qualitative and can predict only the trend rather than actual magnitudes.

\bibliographystyle{unsrtnat}
\bibliography{IzbassarovTammisolaFF10107}
\end{document}